\documentclass[12pt,journal,final,onecolumn]{IEEEtran}

\usepackage[dvips]{graphicx}
\usepackage{epsfig}
\usepackage[cmex10]{amsmath}
\usepackage{amssymb}
\usepackage{amsthm}
\usepackage{amsfonts}
\usepackage{bm}
\usepackage{dsfont}
\usepackage{color}
\usepackage[mathscr]{eucal}
\usepackage{cite}
\usepackage{color}

\graphicspath{{figs/}}

\input{niesen.def}

\newcommand{\RL}{R_{\textup{\tsf{L}}}}

\begin{document}

\bibliographystyle{ieeetr}

\title{The Degrees of Freedom of Compute-and-Forward}

\author{Urs~Niesen, and Phil~Whiting%
\thanks{U. Niesen and P. Whiting are with the
Mathematics of Networks and Communications Research Department, Bell
Labs, Alcatel-Lucent.
Emails: urs.niesen@alcatel-lucent.com, pwhiting@research.bell-labs.com.}%
\thanks{The material in this paper was presented in part at the 2011 IEEE
International Symposium on Information Theory.}%
\thanks{This work was supported in part by AFOSR under grant FA9550-09-1-0317.}%
}

\maketitle

\begin{abstract} 
    We analyze the asymptotic behavior of compute-and-forward relay
    networks in the regime of high signal-to-noise ratios. We consider a
    section of such a network consisting of $K$ transmitters and $K$
    relays. The aim of the relays is to reliably decode an invertible
    function of the messages sent by the transmitters. An upper bound on
    the capacity of this system can be obtained by allowing full
    cooperation among the transmitters and among the relays,
    transforming the network into a $K\times K$ multiple-input
    multiple-output (MIMO) channel. The number of degrees of freedom of
    compute-and-forward is hence at most $K$. In this paper, we analyze
    the degrees of freedom achieved by the lattice coding implementation
    of compute-and-forward proposed recently by Nazer and Gastpar. We
    show that this lattice implementation achieves at most
    $2/(1+1/K)\leq 2$ degrees of freedom, thus exhibiting a very
    different asymptotic behavior than the MIMO upper bound. This raises
    the question if this gap of the lattice implementation to the MIMO
    upper bound is inherent to compute-and-forward in general. We answer
    this question in the negative by proposing a novel
    compute-and-forward implementation achieving $K$ degrees of freedom.
\end{abstract}

\section{Introduction}
\label{sec:intro}

The two central problems of reliable communication over a wireless relay
network are the signal interactions introduced by the wireless medium
and the additive noise experienced at the nodes in the network.
Traditional approaches of dealing with these problems fall broadly into
two categories. On the one hand, intermediate relays in the network can
try to completely remove the receiver noise. The
\emph{decode-and-forward} scheme (see \cite{cover79, laneman04,
kramer05}, among others) falls into this category.  While this solves
the problem of noisy reception, its performance is adversely affected by
the signal interactions, which are usually avoided by careful scheduling
of transmissions. On the other hand, intermediate relays can try to make
use of the signal interactions introduced by the channel either by not
removing the additive noise at all, or by only removing it partially.
Schemes such as \emph{amplify-and-forward} (see, e.g., \cite{schein00,
laneman04, gastpar05, niesen10b}) or \emph{compress-and-forward} (see,
e.g., \cite{cover79, kramer05, kim08, aleksic09, sanderovich09,
avestimehr11}) fall into this category. Since noise is not or only
partially removed at the relays, these schemes suffer from noise
accumulation.

A third approach, referred to as either \emph{compute-and-forward}
\cite{nazer11,nam10}, \emph{physical-layer network coding}
\cite{zhang06,wilson10}, or \emph{analog network coding} \cite{katti07},
aims to both harness the signal interactions introduced by the channel
and remove the noise at the relays. This is achieved by allowing the
relays to decode noiseless \emph{functions} of the transmitted messages.
At the destination node all the information streams are combined 
to determine the original messages being sent. In this paper, we examine
the design and performance of such schemes. 

\begin{figure}[htbp]
    \begin{center}
        \hspace{-0.4cm}\scalebox{0.667}{\input{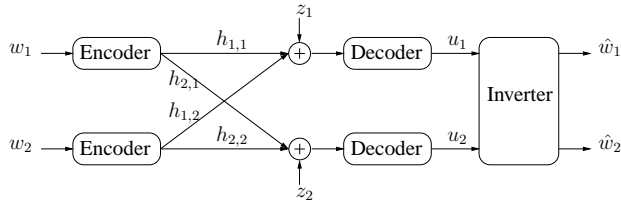}} 
    \end{center}

    \caption{A section of a relay network with two transmitters and two
    relays.}

    \label{fig:intro}
\end{figure}

A small example illustrates the approach, see Fig.~\ref{fig:intro}.
Consider a section of a larger relay network with $2$ transmitters and
$2$ relays. The channel gains $(h_{m,k})$ between the transmitters and
the relays are assumed to be constant and known throughout the network.
The transmitters have access independent messages $w_1, w_2$, which are
separately encoded, modulated, and then sent over the channel.  The
relays receive a linear combination of these transmitted signals
corrupted by additive noise. Each relay decodes independently; however,
the receivers do not aim to decode the original messages $w_1, w_2$.
Rather, each relay $m$ decodes an intermediate quantity $u_m$, which is
a noiseless function of the messages $w_1$ and $w_2$.  Crucially, these
functions are chosen to be adapted to the channel gains. In other words,
the computation of the functions $u_1, u_2$ is aided by the signal
interactions introduced by the channel. Following the decoding stage,
the decoded functions $u_1, u_2$ are combined and inverted to recover
the original messages $w_1, w_2$. This combining and inverting of the
decoded functions is to be interpreted as taking place at the
destination node (not explicitly modeled in this scenario), which is
interested only in the original messages.

In \cite{nazer11}, Nazer and Gastpar propose an ingenious coding scheme
for compute-and-forward using lattice codes (see \cite{loeliger97,
erez04}, among others) at each transmitter. These lattice codes have the
property that any integer linear combination of two codewords is again a
codeword. Due to the additive nature of the channel, each relay receives
a linear combination of the lattice codewords (which is again a
codeword) plus some additive noise. The relays then decode the linear
combination of the codewords, removing the noise. The relays thus decode
a noiseless function of the messages. In terms of our example with two
sources and relays, we see that the decoded quantities $u_1, u_2$ are
linear functions of the messages $w_1, w_2$ in this case. Assuming the
resulting system of linear equations is invertible, the original
messages $w_1$ and $w_2$ can be recovered at the final destination from
$u_1$ and $u_2$.

However, there is a subtle difficulty with this approach that we have
neglected in the above description. The lattice property of the codes
ensures only that any \emph{integer} linear combination of codewords is
again a codeword, whereas the linear combination computed by the
wireless channel can have arbitrary \emph{real} (or complex)
coefficients. To overcome this difficulty, \cite{nazer11} proposes to
scale the received channel output so that the scaled received linear
combination of codewords is close to an integer linear combination. In
general, the larger the scaling factor the better the approximation,
increasing achievable rates. At the same time, a larger scaling factor
results in amplification of noise, decreasing achievable rates.

We hence see that there is a tradeoff between closeness of approximation
and noise amplification. This tradeoff is a central theme in the field
of \emph{Diophantine approximation}, which studies the approximation of
real numbers by rationals, and we will refer to this as the
\emph{Diophantine tradeoff} in compute-and-forward. The rates achievable
by the lattice coding implementation of compute-and-forward in
\cite{nazer11} are not given by an analytic expression, but rather as
the solution to an optimization problem, in which this tradeoff appears
implicitly. It is hence not clear how significant the loss due to this
Diophantine tradeoff is.

In this paper, we show that the loss in rate due to this tradeoff is
indeed significant at high but still realistic values of signal-to-noise
ratio (SNR), say $20$dB and above. In particular, for the two-user
example discussed earlier, we show that due to this Diophantine tradeoff
the compute-and-forward scheme in \cite{nazer11} achieves only one
degree of freedom (capacity pre-log factor), the same as time sharing
between the transmitters. In other words, in the two-user case, the
compute-and-forward implementation in \cite{nazer11} and time sharing
have the same high-SNR behavior. For the general case with $K$
transmitters and $K$ relays, we show that the lattice scheme achieves at
most $2/(1+1/K) \leq 2$ degrees of freedom. While potentially better
than time sharing, this is considerably worse than the MIMO upper bound
of $K$ degrees of freedom that would be achievable with full cooperation
among the transmitters and among the relays.

This negative result raises the question as to whether this Diophantine
tradeoff and the associated loss are inherent to compute-and-forward as
a scheme in general or whether they are an artifact of the
implementation in \cite{nazer11}. We show that the latter is the case
and that compute-and-forward in general does not suffer from this
tradeoff. To this end, we propose a novel implementation of
compute-and-forward that achieves $K$ degrees of freedom, matching the
MIMO upper bound. Thus, compute-and-forward can achieve the same
asymptotic rates as if cooperation among the transmitters and among the
relays were allowed. The proposed achievable scheme introduces the
concept of \emph{signal alignment}, related to the alignment of
\emph{interference}. This alignment of signals is crucial to achieve the
$K$ degrees of freedom upper bound, and indicates that the
compute-and-forward problem and the interference channel problem are
closely related.

The remainder of this paper is organized as follows.
Section~\ref{sec:defs} provides a general formulation of the
compute-and-forward setting.  Section~\ref{sec:main} states the main
results. Proofs are presented in
Sections~\ref{sec:proofspreliminaries}--\ref{sec:proofslattice3}.
Section~\ref{sec:conclusion} contains concluding remarks.

\section{Problem Statement and Notation}
\label{sec:defs}

\subsection{Notational Conventions}
\label{sec:defs_notation}

Throughout this paper, we use the following notational conventions.
Vectors and matrices are written in bold font in lower and upper case,
respectively, e.g., $\bm{h}$ and $\bm{H}$. For a matrix $\bm{H}$, its
transpose is denoted by $\bm{H}^\T$, and its determinant by
$\det(\bm{H})$. For a vector $\bm{h}$, we write $\norm{\bm{h}}$ for its
Euclidean norm. We denote Lebesgue measure by $\mu$. We say that a
property holds for almost every $\bm{H}$ if the set $B$ of $\bm{H}$ for
which the property does \emph{not} hold has Lebesgue measure $\mu(B)$
equal to zero. Finally, all logarithms are to the base $2$, and
therefore channel capacities are expressed in bits per channel use.

\subsection{Problem Statement}
\label{sec:defs_problem}

We consider a section of a relay network with $K$ transmitters and $K$
relays modeled by a discrete-time real Gaussian
channel.\footnote{Throughout this paper, we assume real channels. Using
arguments similar to the ones in~\cite{maddah-ali10,kleinbock04}, the
results can be extended to hold for complex channels as well.} The
\emph{channel output} $y_m[t]$ at receiver $m\in\{1,\ldots,K\}$ and time
$t\in\N$ is 
\begin{equation}
    \label{eq:channel}
    y_m[t] \defeq \sum_{k=1}^K h_{m,k}x_k[t] + z_m[t].
\end{equation}  
Here $x_k[t]\in\R$ is the \emph{channel input} at transmitter
$k\in\{1,\ldots,K\}$, $h_{m,k}\in\R$ is the \emph{channel gain} between
transmitter $k$ and receiver $m$, and $z_m[t]\in\R$ is additive white
Gaussian \emph{noise} with zero mean and unit variance. Note that the
channel gains $(h_{m,k})$ are deterministic and constant across time. As
such, they are known throughout the network.  To simplify notation, let
the row vector 
\begin{equation*}
    \bm{h}_m \defeq 
    \begin{pmatrix}
        h_{m,1} & h_{m,2} & \cdots & h_{m,K}
    \end{pmatrix}
\end{equation*}
be the channel gains to receiver $m$, and set
\begin{equation*}
    \bm{H} \defeq 
    \begin{pmatrix}
        \bm{h}_1 \\
        \bm{h}_2 \\
        \vdots \\
        \bm{h}_K
    \end{pmatrix}.
\end{equation*}

Transmitter $k$ has access to an independent \emph{message} $w_k$
uniformly distributed over $\{0, 1, \ldots, W_k-1\}$.  The goal of
receiver $m$ is to compute the (deterministic) function 
\begin{equation*}
    u_m \defeq a_m(w_1, w_2, \ldots, w_K)\in \{0, 1, \ldots, U_m-1\}.
\end{equation*}
Since $a_m$ is a deterministic function, its range can contain at most
$U_m \leq \prod_{k=1}^K W_k$ elements. We impose that the messages
$(w_k)$ can be recovered from the decoded equations $(u_m)$, i.e.,
that the vector map induced by the $K$ functions $(a_m)$ is
invertible. 

Formally, a \emph{block code} of length $T$ and power constraint $P$
consists of $K$ encoders
\begin{equation*}
    f_k\colon \{0,\ldots, W_k-1\} \to \R^T
\end{equation*}
for $k\in\{1,\ldots, K\}$, mapping the message $w_k$ to channel inputs 
\begin{equation*}
    (x_k[t])_{t=1}^T \defeq f_k(w_k)
\end{equation*}
such that
\begin{equation*}
    \frac{1}{T}\norm{f_k(w_k)}^2 \leq P,
\end{equation*}
and $K$ decoders
\begin{equation*}
    \phi_m\colon \R^T \to \{0, 1, \ldots, U_m-1\}
\end{equation*}
for $m\in\{1,\ldots, K\}$, mapping the channel outputs
$(y_m[t])_{t=1}^T$ to the estimate
\begin{equation*}
    \hat{u}_m \defeq \phi_m\bigl((y_m[t])_{t=1}^T\bigr)
\end{equation*}
of $u_m$.  The \emph{probability of error} of this block code is
\begin{equation*}
    \Pp\bigl( \cup_{m\in\{1,\ldots, K\}} \{ \hat{u}_m \neq u_m \} \bigr).
\end{equation*}
Observe that the probability of error is defined with respect to the
equations $u_m$ and not the original messages $w_k$.  The
(sum) \emph{rate} of this block code is
\begin{equation*}
    \frac{1}{T}\sum_{k=1}^K \log(W_k).
\end{equation*}
A rate $R(\bm{H}, P, (a_m))$ is \emph{achievable} if for every $\eta>
0$ there exists a block code of length $T$ and power constraint $P$ with
probability of error less than $\eta$ and rate at least $R(\bm{H}, P,
(a_m))$. The \emph{computation capacity for functions $(a_m)$},
denoted by $C(\bm{H}, P, (a_m))$, is defined as the supremum of
achievable rates. Finally, define the \emph{computation capacity}
\begin{equation*}
    C(\bm{H}, P) \defeq \sup_{(a_m)} C(\bm{H}, P, (a_m)),
\end{equation*}
where the supremum is over all invertible (deterministic) functions
$(a_m)$. 

Note that in this definition of computation capacity, it is irrelevant
which functions the receivers decode, as long as all the decoded
equations allow recovery of the original messages. This requirement is
best understood in the context of a larger relay network, in which the
channel considered here is only one component of the network, and the
receivers here correspond to intermediate relays. The invertibility of
the map $(a_m)$ guarantees that collectively the decoded equations
$(u_m)$ at these relays contain all the information about the messages
$(w_k)$ at the transmitters. However, the decoded equations have to be
deterministic, i.e., all noise introduced by the channel has to be
removed at the relays. This ensures that noise is not forwarded further
down the larger relay network. These two requirements (invertibility and
noise removal) are the essence of the compute-and-forward approach. We
point out that decode-and-forward is a special case of the above
definition in which the function $a_m$ are given by 
\begin{equation*}
    u_m = a_m(w_1, w_2, \ldots, w_K) = w_m
\end{equation*}
for all $m$. On the other hand, schemes like amplify-and-forward or 
compress-and-forward do not satisfy the above definition, since they
compute randomized (i.e., noisy) functions of the messages. 

While the above definition of computation capacity allows for arbitrary
functions $a_m$ it is worth mentioning the special case of \emph{linear}
functions. In this case, receiver $m$ aims to compute the function 
\begin{equation*}
    u_m \defeq \sum_{k=1}^K a_{m,k}w_k,
\end{equation*}
with $a_{m,k}\in\R$.\footnote{This setting can be slightly generalized
by considering a \emph{vector} of messages $\bm{w}_k$ (instead of a
scalar $w_k$) and computing $\bm{u}_k$ by applying the same linear
function to every component of $\bm{w}_k$. The distinction between the
scalar and vector cases is immaterial for the purpose of this paper, and
we will refer to both as linear computation.} Define the row vector
\begin{equation*}
    \bm{a}_m \defeq 
    \begin{pmatrix}
        a_{m,1} & a_{m,2} & \cdots & a_{m,K}
    \end{pmatrix}
\end{equation*}
and the corresponding matrix
\begin{equation*}
    \bm{A} \defeq 
    \begin{pmatrix}
        \bm{a}_1 \\ 
        \bm{a}_2 \\
        \vdots \\
        \bm{a}_K
    \end{pmatrix}.
\end{equation*}
The messages $(w_k)$ can in this case be recovered from the decoded
equations $(u_m)$ if the matrix $\bm{A}$ is full rank. With slight
abuse of notation, we write $C(\bm{H}, P, \bm{A})$ for the
computation capacity for the linear function determined by the
coefficient matrix $\bm{A}$.

In the remainder of his paper, we will be interested in the
\emph{degrees of freedom} of the computation capacity $C(\bm{H}, P)$
defined as
\begin{equation*}
    \lim_{P\to\infty}\frac{C(\bm{H},P)}{\tfrac{1}{2}\log(P)}
\end{equation*}
assuming the limit exists. If this limit is equal to $D$, then
\begin{equation*}
    C(\bm{H},P) = \frac{D}{2}\log(P)+o(\log(P))
\end{equation*}
as $P\to\infty$. Thus, the degrees of freedom describe the behavior of
$C(\bm{H},P)$ at high SNR. Since the $o(\log(P))$ approximation alone
can be quite weak, we will provide tighter second-order asymptotics as
well.

\section{Main Results}
\label{sec:main}

Nazer and Gastpar \cite[Theorems~1 and 2]{nazer11} provide an achievable
scheme based on lattice codes for computation of linear equations over
the channel \eqref{eq:channel}, showing that, for
$\bm{A}\in\Z^{K\times K}$,
\begin{align}
    \label{eq:rl}
    C(\bm{H}, P,\bm{A})
    & \geq \RL(\bm{H}, P, \bm{A}) \nonumber\\
    & \defeq \sum_{k=1}^K \min_{m: a_{m,k}\neq 0}\RL(\bm{h}_m, P, \bm{a}_m) \nonumber\\
    & \defeq \sum_{k=1}^K \min_{m: a_{m,k}\neq 0}
    \Biggl(
    \frac{1}{2} \log \bigl( 1 + P \norm{\bm{h}_m}^2 \bigr)
    - \frac{1}{2} \log \biggl( 
    \norm{\bm{a}_m}^2 +
    P\Bigl( \norm{\bm{h}_m}^2\norm{\bm{a}_m}^2- \bigl( \bm{h}_m\bm{a}_m^{\T} \bigr)^2 \Bigr)
    \biggr)
    \Biggr)
\end{align}
is achievable. We emphasize that \eqref{eq:rl} is only valid for
\emph{integer} matrices $\bm{A}\in\Z^{K\times K}$. This restriction
turns out to be a significant limitation, as we will see later.

Let us interpret the terms in the definition of
$\RL(\bm{h}_m, P, \bm{a}_m)$. The first term corresponds to the sum
capacity of a multiple-access channel with channel gains $\bm{h}_m$. The
second term represents the rate loss incurred by using the
coefficients $\bm{a}_m$.  This rate loss, governed by
\begin{equation}
    \label{eq:loss}
    \norm{\bm{a}_m}^2+
    P \Bigl( \norm{\bm{h}_m}^2 \norm{\bm{a}_m}^2
    - \bigl( \bm{h}_m \bm{a}_m^{\T} \bigr)^2 \Bigr),
\end{equation}
consists of two parts: the squared norm of $\bm{a}_m$, and the power $P$
times the gap arising from the Cauchy-Schwarz inequality, which is
therefore nonnegative. This second term is zero if and only if
$\bm{a}_m$ and $\bm{h}_m$ are collinear. Recall that $\bm{a}_m$ has
integer components and can therefore not be chosen to be collinear to
$\bm{h}_m$ in general. Denote by
\begin{equation*}
    \RL(\bm{H}, P) 
    \defeq \max_{\bm{A}\in\Z^{K\times K}: 
    \rank(\bm{A})=K} \RL(\bm{H}, P, \bm{A})
\end{equation*}
the largest rate achievable with the lattice scheme proposed in
\cite{nazer11}.\footnote{By \cite[Lemma~1]{nazer11}, a maximizing
$\bm{A}\in\Z^{K\times K}$ exists.}

As mentioned earlier, the scheme by Nazer and Gastpar uses lattice
codes, which have the property that every integer linear combination of
two codewords is again a codeword. With this approach, the receivers 
directly decode the linear combinations $(u_m)$ and never explicitly
decode the messages $(w_k)$. A different approach would be to choose
$\bm{A}=\bm{I}$ so that $u_m=w_m$ for all $m$. This can be implemented
by time sharing between all the transmitters, achieving a sum rate of at
least
\begin{align}
    \label{eq:ts}
    C(\bm{H}, P) 
    & \geq C(\bm{H}, P, \bm{I}) \nonumber\\
    & \geq \sum_{k=1}^K \frac{1}{2K}\log\bigl(1+KP\abs{h_{k,k}}^2\bigr).
\end{align}
For $\bm{A}=\bm{I}$, the problem actually reduces to the standard
interference channel, for which interference alignment achieves
\begin{align}
    \label{eq:ia}
    C(\bm{H}, P) 
    & \geq C(\bm{H}, P, \bm{I}) \nonumber\\
    & \geq \frac{K}{4}\log(P) - o(\log(P))
\end{align}
as $P\to\infty$ for almost every channel matrix $\bm{H}$
\cite{motahari09}. As $P\to\infty$, this rate is the best achievable for
$\bm{A}=\bm{I}$ and almost every $\bm{H}$, as it is shown in
\cite{host-madsen05} that
\begin{equation}
    \label{eq:ia_upper}
    C(\bm{H}, P, \bm{I})
    \leq \frac{K}{4}\log(P) + o(\log(P)).
\end{equation}

Finally, by allowing cooperation among the transmitters and among the
receivers, the computation rate can be upper bounded by the capacity of
the MIMO channel with the same channel matrix $\bm{H}$. This can be
further upper bounded by relaxing the per-antenna power constraint to a
sum power constraint, so that, by \cite{telatar99},
\begin{align}
    \label{eq:upper}
    C(\bm{H}, P) 
    & \leq \max
    \frac{1}{2}\log\det\bigl(\bm{I}+\bm{H}\bm{Q}\bm{H}^\T\bigr),
\end{align}
where the maximization is over all covariance matrices $\bm{Q}$ with
trace at most $KP$. 

To compare the upper bound \eqref{eq:upper} to the lower bounds
\eqref{eq:rl}, \eqref{eq:ts}, and \eqref{eq:ia}, it is insightful to
consider their asymptotic behavior as power $P$ grows. The time-sharing
lower bound \eqref{eq:ts} yields
\begin{equation*}
    \liminf_{P\to\infty} \frac{C(\bm{H}, P)}{\tfrac{1}{2}\log(P)}
    \geq 1,
\end{equation*}
i.e., time sharing achieves one degree of freedom. The
interference-alignment lower bound \eqref{eq:ia} yields
\begin{equation*}
    \liminf_{P\to\infty} \frac{C(\bm{H}, P)}{\tfrac{1}{2}\log(P)}
    \geq K/2
\end{equation*}
for almost every channel matrix $\bm{H}$, i.e., interference
alignment achieves $K/2$ degrees of freedom. On the other hand, almost
every channel matrix $\bm{H}$ has full rank, in which case the MIMO
upper bound \eqref{eq:upper} yields
\begin{equation}
    \label{eq:upper2}
    \limsup_{P\to\infty} \frac{C(\bm{H}, P)}{\tfrac{1}{2}\log(P)}
    \leq K,
\end{equation}
i.e., the corresponding MIMO channel has $K$ degrees of freedom. Thus,
at high SNRs, time sharing and interference alignment behave very
differently from the MIMO upper bound for almost every $\bm{H}$. Observe
that by \eqref{eq:ia_upper} any scheme using decode-and-forward, i.e.,
with coefficient matrix $\bm{A}=\bm{I}$, achieves at most $K/2$ degrees
of freedom for almost every $\bm{H}$. Hence, if we are to attain the
upper bound of $K$ on the degrees of freedom, the use of general
compute-and-forward (as opposed to simple decode-and-forward) will be
necessary.

The behavior of the rate $\RL$ achieved by the lattice scheme is
more difficult to evaluate. If $\bm{H}\in\Z^{K\times K}$ has integer
components and is invertible, we can set $\bm{A}=\bm{H}$ 
in~\eqref{eq:rl} to obtain
\begin{equation*}
    \RL(\bm{h}_m, P, \bm{h}_m) 
    = \frac{1}{2}\log\bigl(1+P
    \norm{\bm{h}_m}^2\bigr)-\frac{1}{2}\log\bigl(\norm{\bm{h}_m}^2\bigr).
\end{equation*}
Hence, in this case,
\begin{equation*}
    \liminf_{P\to\infty} \frac{\RL(\bm{H}, P)}{\tfrac{1}{2}\log(P)}
    \geq K.
\end{equation*}
More generally, if $\bm{H}\in\Q^{K\times K}$ has rational components and
is invertible, then there exists a $q\in\N$ such that
$q\bm{H}\in\Z^{K\times K}$.  Setting $\bm{A}=q\bm{H}$ in~\eqref{eq:rl}
yields that lattice coding achieves a rate of 
\begin{equation*}
    \RL(\bm{h}_m, P, q\bm{h}_m) 
    = \frac{1}{2}\log\bigl(1+P
    \norm{\bm{h}_m}^2\bigr)-\frac{1}{2}\log\bigl(q^2\norm{\bm{h}_m}^2\bigr),
\end{equation*}
and again
\begin{equation*}
    \liminf_{P\to\infty} \frac{\RL(\bm{H}, P)}{\tfrac{1}{2}\log(P)}
    \geq K.
\end{equation*}
Since $\RL \leq C$, we obtain together with the MIMO upper bound
\eqref{eq:upper} that for invertible $\bm{H}\in\Q^{K\times K}$
\begin{equation*}
    \lim_{P\to\infty} \frac{\RL(\bm{H}, P)}{\tfrac{1}{2}\log(P)}
    = K.
\end{equation*}
In other words, for invertible $\bm{H}$ with \emph{rational} components,
the scheme based on lattice coding is asymptotically optimal. In
particular, this implies that the lattice scheme significantly
outperforms the schemes based on time sharing and based on interference
alignment.

However, the requirement of rational channel gains $\bm{H}$ is quite
strong. In fact, this event has Lebesgue measure zero. The question
arises whether the behavior of the rate $\RL$ achieved by the lattice
scheme of \cite{nazer11} is significantly altered if we relax this
assumption of rational channel gains. The next theorem shows that this
is indeed the case. In fact, for almost all channel gains, the lattice
scheme has an asymptotic behavior that is not significantly better than
time sharing.

\begin{theorem}
    \label{thm:lattice1}
    For any $K\geq 2$ and almost every $\bm{H}\in\R^{K\times K}$ there
    exists a positive constant $c_1=c_1(K,\bm{H})$ such that for all
    $P\geq 3$
    \begin{equation*}
        \RL(\bm{H}, P) \leq \frac{1}{1+1/K}\log(P)+c_1\log\log(P).
    \end{equation*}    
    In particular, this implies that for any $K\geq 2$ and almost every
    $\bm{H}\in\R^{K\times K}$
    \begin{equation*}
        \limsup_{P\to\infty} 
        \frac{\RL(\bm{H}, P)}{\tfrac{1}{2}\log(P)}
        \leq \frac{2}{1+1/K}.
    \end{equation*}
\end{theorem}
We remark that, for $K=2$, Theorem~\ref{thm:lattice1} can be sharpened
to 
\begin{equation}
    \label{eq:k2}
    \limsup_{P\to\infty} \max_{\bm{a}_m\in\Z^2\setminus\{\bm{0}\}}
    \frac{\RL(\bm{h}_m, P, \bm{a}_m)}{\tfrac{1}{2}\log(P)}
    \leq 1/2,
\end{equation}
for almost every $\bm{h}_m\in\R^2$, so that 
\begin{equation*}
    \limsup_{P\to\infty} 
    \frac{\RL(\bm{H}, P)}{\tfrac{1}{2}\log(P)}
    \leq 1
\end{equation*}
for almost every $\bm{H}\in\R^{2\times 2}$.

Theorem~\ref{thm:lattice1} shows that for almost every channel matrix
$\bm{H}$ there is only limited asymptotic gain over time sharing by
using the lattice scheme in \cite{nazer11}. In particular, for $K=2$,
time sharing and lattice coding achieve the same degrees of freedom. For
large $K$, the upper bound in Theorem~\ref{thm:lattice1} is
approximately $2$---better than time sharing, but still far off from the
$K/2$ degrees of freedoms achievable with interference alignment and the
MIMO upper bound of $K$ degrees of freedom. In other words, it seems to
suggest that, at high SNR, compute-and-forward offers only limited
advantage over standard coding schemes. This conclusion turns out to be
misleading, as we will see later.

The bad asymptotic performance of the lattice scheme is due to the rate
loss term \eqref{eq:loss}. As pointed out earlier, to make the second
term in \eqref{eq:loss} small, the coefficients $\bm{a}_m$ should be as
close to collinear to the channel gains $\bm{h}_m$ as possible.
However, since $\bm{a}_m$ is forced to be an integer vector, and since
$\bm{h}_m$ is a real vector, this is in general only possible by
increasing the norm of $\bm{a}_m$. This, in turn, increases the first
term in \eqref{eq:loss}. The tradeoff between the two terms in
\eqref{eq:loss} is a main theme in the field of \emph{Diophantine
approximation}.  In particular, the proof of Theorem~\ref{thm:lattice1}
builds on a result of Khinchin to show that, for almost every channel
gain $\bm{h}_m$, the coefficient vector $\bm{a}_m$ can only be close to
collinear to $\bm{h}_m$ if $\norm{\bm{a}_m}$ is large.

\begin{example}
    \label{eg:lattice}
    Consider the channel vector $\bm{h}= (1 \ \ h_2)$ to one of the
    receiver. Consider
    \begin{equation*}
        \max_{\bm{a}\in\Z^2\setminus\{\bm{0}\}}\RL(\bm{h},P,\bm{a}),
    \end{equation*}
    the maximal rate at which \emph{any} (nontrivial) integer linear
    equation can be decoded at the receiver. From \eqref{eq:k2}, we know
    that
    \begin{equation*}
        \limsup_{P\to\infty}
        \max_{\bm{a}\in\Z^2\setminus\{\bm{0}\}}
        \frac{\RL(\bm{h}, P, \bm{a})}
        {\frac{1}{2}\log(P)} 
        \leq 1/2
    \end{equation*}
    for almost every\footnote{While \eqref{eq:k2} is stated for
    almost every $\bm{h}\in\R^2$, the same arguments can be used to show
    that \eqref{eq:k2} also holds for almost every $\bm{h}$ of the form
    $(1 \ \ h_2)$.} $h_2\in\R$. On the other hand, for $h_2\in\Q$, 
    \begin{equation*}
        \lim_{P\to\infty}
        \max_{\bm{a}\in\Z^2\setminus\{\bm{0}\}}
        \frac{\RL(\bm{h}, P, \bm{a})}
        {\frac{1}{2}\log(P)} 
        = 1.
    \end{equation*}
    While these statements are only valid asymptotically as
    $P\to\infty$, this qualitative behavior is already visible at moderate
    values of SNR, as is depicted in Fig.~\ref{fig:layers}.
    \begin{figure}[!ht]
        \begin{center}
            \scalebox{0.8}{\hspace{-0.4cm}\input{figs/example_lattice_20db.tex}}\\
            \scalebox{0.8}{\hspace{-0.4cm}\input{figs/example_lattice_30db.tex}}\\
            \scalebox{0.8}{\hspace{-0.4cm}\input{figs/example_lattice_40db.tex}}\\
            \scalebox{0.8}{\hspace{-0.4cm}\input{figs/example_lattice_50db.tex}}
        \end{center}

        \caption{Normalized rate $\max_{\bm{a}}\RL(\bm{h}, P,
        \bm{a})/\tfrac{1}{2}\log(1+\bm{h}^2P)$ achievable with lattice
        codes \cite{nazer11} with optimized coefficient vector
        $\bm{a}\in\Z^2\setminus\{\bm{0}\}$ for channel gain $\bm{h} = (1
        \ \  h_2)$ as a function of $h_2\in[0,1]$. The plots are for a
        value $P$ of $20$dB, $30$dB, $40$dB, and $50$dB (from top to
        bottom). As $P\to\infty$, the normalized rate converges to at
        most $1/2$ for almost every value of $h_2$. On the other hand,
        for $h_2\in\Q$ (a set of measure zero), the normalized rate
        converges to $1$.  This limiting behavior can already be
        observed at the values of SNR shown here.} 
        
        \label{fig:layers}
    \end{figure}
\end{example}

We now introduce a different implementation of the compute-and-forward
approach that achieves $K$ degrees of freedom, matching the asymptotic
behavior of the MIMO upper bound \eqref{eq:upper2}. In other words, even
though both the receivers and the transmitters are distributed, the
proposed communication scheme achieves the same number of degrees of
freedom as a centralized communication scheme in which all transmitters can
cooperate and all receivers can cooperate.

\begin{theorem}
    \label{thm:lattice3}
    For every $K\geq 2$ and almost every $\bm{H}\in\R^{K\times K}$
    there exist positive constants $c_2=c_2(K,\bm{H})$ and
    $c_3=c_3(K,\bm{H})$ such that for all $P\geq 2$ 
    \begin{equation*}
        \frac{K}{2}\log(P)
        - c_2\log^{\tfrac{K^2}{1+K^2}}(P)
        \leq C(\bm{H},P) 
        \leq \frac{K}{2}\log(P)+c_3.
    \end{equation*}
    In particular, this implies that for any $K\geq 2$ and almost every
    $\bm{H}\in\R^{K\times K}$,
    \begin{equation*}
        \lim_{P\to\infty} 
        \frac{C(\bm{H}, P)}{\tfrac{1}{2}\log(P)}
        = K.
    \end{equation*}
\end{theorem}

Recall that the implementation of compute-and-forward in \cite{nazer11}
uses lattice/linear codes together with output scaling. The aim of this
output scaling is to make the scaled channel gains close to integer. The
difficulty with this approach is that the scaling of the channel outputs
amplifies the additive receiver noise. In order for the scaled channel
gains to be close to integer, the scaling factor should be large. On the
other hand, in order to have small noise amplification, the scaling
factor should be small. These two conflicting requirements result in the
Diophantine tradeoff mentioned in the introduction. This tradeoff can be
observed in the tension between the two terms in \eqref{eq:loss} as
discussed earlier.

Our proposed achievable scheme in Theorem~\ref{thm:lattice3} also uses
linear codes at the transmitters. However, it avoids the scaling of the
channel outputs and thereby the Diophantine tradeoff. Instead, we use a
modulation scheme based on \emph{signal alignment} over the real numbers
to convert the real linear combinations produced by the channel into
integer linear combinations. This step builds on a construction
suggested recently for the alignment of \emph{interference} in
\cite{motahari09,etkin09}, which itself is based on prior work on
Diophantine approximation on manifolds \cite{bernik01, beresnevich02}.
The proposed approach is best illustrated with an example.

\begin{example}
    Consider again the $K=2$ case. Assume the channel gains are of the
    form
    \begin{equation*}
        \bm{H} \defeq
        \begin{pmatrix}
            1 & h_2 \\
            h_1 & 1
        \end{pmatrix}.
    \end{equation*}

    Set the channel input to be
    \begin{align*}
        x_1 & \defeq \bar{w}_1,\\
        x_2 & \defeq \bar{w}_2,
    \end{align*}
    where both $\bar{w}_k$ are codewords from the same lattice code. The
    channel output is then
    \begin{align*}
        y_1 & = \bar{w}_1+h_2\bar{w}_2+z_1,\\
        x_2 & = h_1\bar{w}_1+\bar{w}_2+z_2.
    \end{align*}
    Given that both codewords are from the same lattice code, one might
    hope that an integer combination of them might be decodable at
    higher rates than the individual messages themselves. However, the
    arguments in Theorem~\ref{thm:lattice1} show that, for almost all
    $\bm{H}$ and at high enough SNR, each receiver can essentially
    decode both $\bar{w}_1$ and $\bar{w}_2$ whenever it can decode an
    integer combination of them. This limits the computation rate to one
    degree of freedom.

    A simple improvement over this scheme is to set
    \begin{align*}
        x_1 & \defeq \bar{w}_1,\\
        x_2 & \defeq h_1\bar{w}_2,
    \end{align*}
    The channel output is now
    \begin{align*}
        y_1 & = \bar{w}_1+h_1h_2\bar{w}_2+z_1,\\
        x_2 & = h_1(\bar{w}_1+\bar{w}_2)+z_2.
    \end{align*}
    This results in the signals $\bar{w}_1$ and $\bar{w}_2$ to be both
    observed with the same effective channel gain $h_1$ at receiver two.
    In other words, we have signal alignment at the second receiver.
    However, the signals at the first receiver are still unaligned. This
    limits the computation rate to again only one degree of freedom.

    To achieve alignment at both receivers, split the messages into two
    parts, and set
    \begin{align*}
        x_1 & \defeq \bar{w}_{1,1}+h_1h_2\bar{w}_{1,2},\\
        x_2 & \defeq h_1\bar{w}_{2,1}+h_1^2h_2\bar{w}_{2,2}.
    \end{align*}
    This results in the channel outputs
    \begin{align*}
        y_1 & = \bar{w}_{1,1}
        +h_1h_2(\bar{w}_{1,2}+\bar{w}_{2,1})
        +h_1^2h_2^2\bar{w}_{2,2}+z_1,\\
        x_2 & = h_1(\bar{w}_{1,1} +\bar{w}_{2,1})
        +h_1^2h_2(\bar{w}_{1,2}+\bar{w}_{2,2})+z_2.
    \end{align*}
    We now have partial alignment at both receivers. Receiver one
    decodes $\bar{w}_{1,1}$, $\bar{w}_{1,2}+\bar{w}_{2,1}$, and
    $\bar{w}_{2,2}$. Receiver two decodes $\bar{w}_{1,1} +\bar{w}_{2,1}$
    and $\bar{w}_{1,2}+\bar{w}_{2,2}$. It can be shown that this
    achieves a computation rate of $4/3$ degrees of freedom. 

    By breaking the messages into more submessages and aligning them
    pairwise in the same manner, this construction achieves a
    computation rate approaching two degrees of freedom, as promised in
    Theorem~\ref{thm:lattice3}.
\end{example}

\begin{remark}
    In the channel model \eqref{eq:channel}, the channel gains $\bm{H}$
    are assumed to be constant and as such known everywhere. In
    practice, this channel state information (CSI) would have to be
    estimated and distributed throughout the network, resulting in
    signaling overhead. 
    
    Having access to CSI at both the receivers and the transmitters is
    critical for the operation of the compute-and-forward scheme
    proposed here (achieving $K$ degrees of freedom) as well as for the
    interference alignment scheme in \cite{motahari09} (achieving $K/2$
    degrees of freedom). In contrast, the lattice coding implementation
    of compute-and-forward in \cite{nazer11} (and shown here to achieve
    at most $2$ degrees of freedom) requires only CSI at the receivers
    but not at the transmitters. Whether the lattice coding scheme in
    \cite{nazer11} achieves the optimal degrees of freedom if the use of
    transmitter CSI is excluded is an open question.
\end{remark}

\begin{remark}
    Throughout this paper, we have been concerned almost exclusively
    with degrees of freedom. The second-order asymptotics in
    Theorem~\ref{thm:lattice3} are quite poor, especially for larger
    values of $K$. Deriving tighter approximations valid for moderate
    values of SNR is an interesting direction for further investigation. 
\end{remark}

\section{Preliminaries: Diophantine Approximation}
\label{sec:proofspreliminaries}

In all of the proofs, we will be using facts from Diophantine
approximation. Here we provide the necessary background as well as some
extensions of well-known results.

Let $h$ be a real and $a,q$ be integers. How well can $h$ be
approximated by the ratio $a/q$? Since the rationals $\Q$ are dense in
the reals $\R$, this can be done to any arbitrary degree of accuracy.
However, to get a good approximation, the denominator $q$ will, in
general, have to be large. The question then becomes one of quantifying
the tradeoff between the quality of approximation and the size of $q$.
Formally, the problem is to analyze the behavior of
\begin{equation}
    \label{eq:prelim_khinchin}
    \min_{a\in\Z}\abs{h-a/q} 
\end{equation}
as a function of $q\in\N$ for fixed $h\in\R$. A result due to Khinchin
(see, e.g., \cite[Theorem~1]{sprindzuk79}) states that if $\psi$ is a
nonnegative function such that
\begin{equation}
    \label{eq:prelim_khinchin2}
    \sum_{q=1}^\infty q \psi(q) 
\end{equation}
\emph{converges}, then for almost every $h\in\R$ there exists a positive
constant $c=c(h)$ such that
\begin{equation*}
    \min_{a\in\Z}\abs{h-a/q} \geq c \psi(q)
\end{equation*}
for all $q\in\N$. On the other hand, if \eqref{eq:prelim_khinchin2}
\emph{diverges}, then for almost every $h\in\R$ and every positive
constant $c$, there are infinitely many values of $q\in\N$ such that
\begin{equation*}
    \min_{a\in\Z}\abs{h-a/q} \leq c \psi(q)
\end{equation*}
The convergent and divergent parts of Khinchin's theorem show that for
almost every $h\in\R$ the approximation error $\abs{h-a/q}$ can be made
to decay at least as fast as $O(q^{-2+\delta})$ but no faster than
$\Omega(q^{-2-\delta})$ for any $\delta >0$.  

The next lemma provides a simple generalization of the convergent part
of Khinchin's theorem to more than one dimension and to approximations
with denominator $\sqrt{q}$.

\begin{lemma}
    \label{thm:khinchin1}
    Let $\psi\colon \N\to\Rp$. If 
    \begin{equation*}
        \sum_{q=1}^\infty (\sqrt{q}\psi(q))^K 
        < \infty,
    \end{equation*}
    then for almost every $\bm{h}\in\R^K$ there is a positive constant
    $c=c(K,\bm{h})$ such that 
    \begin{equation*}
        \max_{k\in\{1, \ldots, K\}}\min_{a_k\in\Z}\abs{h_k-a_k/\sqrt{q}} 
        \geq c\psi(q)
    \end{equation*}
    for all $q\in\N$.
\end{lemma}

The lemma implies that, for almost every $\bm{h}\in\R^K$, the
approximation error 
\begin{equation*}
    \max_{k}\min_{a_k\in\Z}\abs{h_k-a_k/\sqrt{q}}
\end{equation*}
can decay no faster than $\Omega(q^{-1/2-1/K-\delta})$ for
any $\delta > 0$. 

\begin{IEEEproof}
    Let $B_q$ be the vectors $\bm{h}\in [0,1)^K$ such that
    \begin{equation}
        \label{eq:thm_khinchin1}
        \max_{k}\min_{a_k\in\Z}\abs{h_k-a_k/\sqrt{q}} 
        \leq \psi(q).
    \end{equation}
    Since $h_k\in[0,1)$, the integer $a_k$ can be restricted to the set
    $\{0,\ldots, \ceil{\sqrt{q}}\}$ for all $k$. Setting $\bm{a}=(a_k)$,
    we see that a vector $\bm{h}$ is in $B_q$ if and only if it is at a
    $\ell_\infty$ distance of at most $\psi(q)$ of such a vector
    $\bm{a}/\sqrt{q}$. Thus, each $\bm{a}\in\{0,\ldots,
    \ceil{\sqrt{q}}\}^K$ contributes at most a subset of volume
    $(2\psi(q))^K$ to $B_q$. Since there are at most $(\sqrt{q}+2)^K$
    such vectors $\bm{a}$, we have
    \begin{equation*}
        \mu(B_q) \leq (\sqrt{q}+2)^K (2\psi(q))^K.
    \end{equation*}
    By the convergence assumption, this implies that
    \begin{equation*}
        \sum_{q=1}^\infty \mu(B_q) 
        \leq \sum_{q=1}^\infty (\sqrt{q}+2)^K (2\psi(q))^K 
        < \infty.
    \end{equation*}
    Applying the Borel-Cantelli lemma (see, e.g.,
    \cite[Theorem~1.6.1]{durret04}), this shows that
    \begin{equation*}
        \mu(B_q \text{ i.o.}) = 0,
    \end{equation*}
    where ``i.o.'' stands for ``infinitely often'' (as a function of
    $q$). Thus, almost every $\bm{h}\in[0,1)^K$ satisfies
    \eqref{eq:thm_khinchin1} only finitely many times. Since $\R^K$ is
    the countable union of integer cubes $\prod_{k=1}^K [b_k, b_k+1)$,
    the same holds also for almost all $\bm{h}\in\R^K$. 

    Fix a $\bm{h}\in\R^K$ for which \eqref{eq:thm_khinchin1} holds only
    finitely many times. Then there exists a finite number $Q(\bm{h})$
    such that
    \begin{equation*}
        \max_{k}\min_{a_k\in\Z}\abs{h_k-a_k/\sqrt{q}} 
        \geq \psi(q)
    \end{equation*}
    for all $q\geq Q(\bm{h})$. Set
    \begin{align*}
        c = c(K,\bm{h}) 
        \defeq \min\Biggl\{1,\min_{q\in\{1,\ldots, Q(h)\}}
        \frac{\displaystyle \max_{k}\min_{a_k\in\Z}
        \abs{h_k-a_k/\sqrt{q}}}{\psi(q)}\Biggr\},
    \end{align*}
    and observe that $c$ is positive. Then
    \begin{equation*}
        \max_{k}\min_{a_k\in\Z}\abs{h_k-a_k/\sqrt{q}} 
        \geq c\psi(q)
    \end{equation*}
    for all $q\in\N$, concluding the proof of the lemma.
\end{IEEEproof}

We will also need a generalization of the convergent part of Khinchin's
theorem to manifolds in Euclidean space. We start with a small example
to illustrate the setting. Consider again the question of rational
approximation in \eqref{eq:prelim_khinchin}. 
This can be generalized to several dimensions as follows. Fix 
$\bm{H}\in\R^{K\times K}$; what is the behavior of
\begin{equation*}
    \min_{a\in\Z}\big\lvert {\textstyle\sum_{k,m}} q_{m,k} h_{m,k} - a \big\rvert 
\end{equation*}
as a function of $q_{m,k}\in\Z$? The generalization of Khinchin's
theorem to this setting is referred to as Groshev's theorem. 

A further generalization, and the one that will be needed in this paper,
is to allow for \emph{functions} of $\bm{H}$. Let $\mc{G}$ be a
collection of functions $g\colon \R^{K\times K}\to\R$. Fix
$\bm{H}\in\R^{K\times K}$; what is the behavior of
\begin{equation*}
    \min_{a\in\Z}\big\lvert {\textstyle\sum_{g\in\mc{G}}} q_{g} g(\bm{H}) - a \big\rvert 
\end{equation*}
as a function of $q_{g}\in\Z$?  In particular, we will be interested in
the collection of functions
\begin{equation}
    \label{eq:mcgl}
    \mc{G}_L 
    \defeq \Biggl\{
    \prod_{k=1}^K\prod_{m=1}^K
    h_{m,k}^{s_{m,k}}:
    \bm{S}\in\{0,\ldots, L-1\}^{K\times K} 
    \Biggr\}.
\end{equation}
In words, $\mc{G}_L$ is the collection of all monomials in the channel
gains $\bm{H}$ with exponents between $0$ and $L-1$. In the following,
we will usually fix a particular realization of $\bm{H}$ and treat the
set $\mc{G}_L$ as a collection of $L^{K^2}$ points in $\R$.

\begin{remark}
    \label{rem:unique}
    It is straightforward to verify that, for almost every
    $\bm{H}\in\R^{K\times K}$, all $L^{K^2}$ monomials in $\mc{G}_L$
    evaluate to distinct numbers. This implies that, for almost every
    $\bm{H}\in\R^{K\times K}$, we have $\card{\mc{G}_L}=L^{K^2}$.
    Furthermore, again for almost all $\bm{H}$, every $g\in\mc{G}_L$ can
    be uniquely factorized into powers of $h_{m,k}$. In other words, to
    each $g$ corresponds a \emph{unique} set of powers $\bm{S}$ such
    that
    \begin{equation*}
        g = {\textstyle\prod_{m,k}}h_{m,k}^{s_{m,k}}.
    \end{equation*}
    We refer to this as the \emph{unique factorization} property.  Given
    that we are only interested in results that hold for almost every
    channel matrix $\bm{H}$, we may assume in the following that
    $\mc{G}_L$ has this unique factorization property.
\end{remark}

The following lemma is a special case of a more general result from
\cite{bernik01, beresnevich02} (see also \cite{motahari09}).

\begin{lemma}
    \label{thm:groshev2}
    Let $\psi\colon \N\to\Rp$ be a monotonically decreasing function and
    $L\in\N$, $L\geq 2$. If 
    \begin{equation*}
        \sum_{q=1}^\infty q^{\card{\mc{G}_L}-2}\psi(q)
        < \infty,
    \end{equation*}
    then for almost every $\bm{H}\in\R^{K\times K}$ there is a positive
    constant $c=c(K,\bm{H})$ such that
    \begin{equation*}
        \min_{a\in\Z}
        \big\lvert 
        {\textstyle\sum_{g\in\mc{G}_L, g\neq 1}} q_g g -a\big\rvert
        \geq c \psi\bigl(
        {\textstyle\max_{g\in\mc{G}_L, g\neq 1}} \abs{q_g} 
        \bigr)
    \end{equation*}
    for all $(q_g)\in\Z^{\card{\mc{G}_L}-1}\setminus\{\bm{0}\}$.
\end{lemma}

Lemma~\ref{thm:groshev2} implies that, for almost every
$\bm{H}\in\R^{K\times K}$, the approximation error 
\begin{equation*}
    \min_{a\in\Z}\big\lvert {\textstyle\sum_{g\in\mc{G}_L, g\neq 1}} q_g g -a\big\rvert
\end{equation*}
can decay no faster than 
\begin{equation*}
    \Omega\Bigl(\bigl( {\textstyle\max_{g\in\mc{G}_L, g\neq 1}} 
    \abs{q_g}\bigr)^{-\card{\mc{G}_L}+1-\delta}\Bigr)
\end{equation*}
for any $\delta > 0$.

\section{Proof of Theorem \ref{thm:lattice1}}
\label{sec:proofslattice1}

We want to upper bound the largest rate
\begin{equation}
    \label{eq:rl2}
    \RL(\bm{H},P)
    \defeq \max_{\bm{A}\in\Z^{K\times K}: \rank(\bm{A})=K}\RL(\bm{H}, P, \bm{A}) 
\end{equation}
achievable with the lattice coding scheme of~\cite{nazer11}.  From the
above definition, we see that the coefficient matrix $\bm{A}$ can be
chosen as a function of $P$. In particular, to each $P$ corresponds an
optimal $\bm{A}=\bm{A}(P)$ maximizing the right-hand side of
\eqref{eq:rl2}.\footnote{It can be shown that such an optimal $\bm{A}$
exists, see \cite[Lemma~1]{nazer11}. If more than one maximizer exists,
we choose one of them.} We consider this $\bm{A}$ in the following so
that
\begin{equation}
    \label{eq:rl3}
    \RL(\bm{H},P) = \RL(\bm{H}, P, \bm{A}(P)).
\end{equation}

Recall that
\begin{equation}
    \label{eq:rl4}
    \RL(\bm{H}, P, \bm{A}) 
    = \sum_{k=1}^K \min_{m: a_{m,k}\neq 0}\RL(\bm{h}_m, P, \bm{a}_m),
\end{equation}
and that $\RL(\bm{h}_m, P, \bm{a}_m)$ consists of two terms,
the desired term
\begin{equation*}
    \frac{1}{2} \log \bigl( 1 + P \norm{\bm{h}_m}^2 \bigr) 
\end{equation*}
and the loss term
\begin{equation*}
     - \frac{1}{2} \log \biggl( 
    \norm{\bm{a}_m}^2 +
    P\Bigl( \norm{\bm{h}_m}^2\norm{\bm{a}_m}^2- \bigl( \bm{h}_m\bm{a}_m^{\T} \bigr)^2 \Bigr)
    \biggr).
\end{equation*}
We start by upper bounding $\RL(\bm{h}_m, P, \bm{a}_m)$ for a fixed
value of $m\in\{1,\ldots, K\}$. Together with \eqref{eq:rl4}, this
yields an upper bound on $\RL(\bm{H}, P, \bm{A})$ and hence on
$\RL(\bm{H}, P)$. 

We can rewrite the quantity inside the logarithm of the loss term in
$\RL(\bm{h}_m, P, \bm{a}_m)$ as
\begin{align*}
    \norm{\bm{a}_m}^2 + P\Bigl(\norm{\bm{h}_m}^2\norm{\bm{a}_m}^2 
    - \bigl( \bm{h}_m\bm{a}_m^{\T} \bigr)^2\Bigr)
    & = \norm{\bm{a}_m}^2 
    + P\norm{\bm{h}_m}^2\norm{\bm{a}_m}^2\bigl(1-\cos^2(\angle(\bm{h}_m,\bm{a}_m))\bigr) \\
    & = \norm{\bm{a}_m}^2
    +P\norm{\bm{h}_m}^2\norm{\bm{a}_m}^2\sin^2(\angle(\bm{h}_m,\bm{a}_m)).
\end{align*}
Now, for $x\in[-\pi/2,\pi/2]$, 
\begin{equation*}
    \sin^2(x) \geq \frac{4}{\pi^2}x^2,
\end{equation*}
and, for $\angle(\bm{h}_m,\bm{a}_m)$ measured in $\in[-\pi,\pi]$,
\begin{equation*}
    \abs{\angle(\bm{h}_m,\bm{a}_m)}
    \geq \bigg\lVert\frac{\bm{h}_m}{\norm{\bm{h}_m}}
    -\frac{\bm{a}_m}{\norm{\bm{a}_m}}\bigg\rVert
\end{equation*}
by lower bounding the distance along the great circle on the unit sphere
by its chordal distance.  Since $\RL(\bm{h}_m,P,\bm{a}_m)$ is invariant
to multiplication of $\bm{a}_m$ by $-1$, we can assume that
$\angle(\bm{h}_m,\bm{a}_m)\in[-\pi/2,\pi/2]$ so that
\begin{align}
    \label{eq:thm_lattice1_0}
    \norm{\bm{a}_m}^2 + P\Bigl(\norm{\bm{h}_m}^2\norm{\bm{a}_m}^2 
    - \bigl( \bm{h}_m\bm{a}_m^{\T} \bigr)^2\Bigr) 
    & \geq \norm{\bm{a}_m}^2
    +\frac{4}{\pi^2}P\norm{\bm{h}_m}^2\norm{\bm{a}_m}^2
    \bigg\lVert\frac{\bm{h}_m}{\norm{\bm{h}_m}}
    -\frac{\bm{a}_m}{\norm{\bm{a}_m}}\bigg\rVert^2 \nonumber\\
    & \geq \norm{\bm{a}_m}^2
    +\frac{4}{\pi^2}P\norm{\bm{h}_m}^2\norm{\bm{a}_m}^2
    \max_{1\leq k \leq K}
    \bigg\lvert\frac{h_{m,k}}{\norm{\bm{h}_m}}
    -\frac{a_{m,k}}{\norm{\bm{a}_m}}\bigg\rvert^2 \nonumber\\
    & \geq \norm{\bm{a}_m}^2
    +\frac{4}{\pi^2}P\norm{\bm{h}_m}^2\norm{\bm{a}_m}^2
    \max_{1\leq k \leq K-1}
    \bigg\lvert\frac{h_{m,k}}{\norm{\bm{h}_m}}
    -\frac{a_{m,k}}{\norm{\bm{a}_m}}\bigg\rvert^2\!\!.
\end{align}
As we will see shortly, the restriction of the maximum in the last
inequality to $k\in\{1,\ldots, K-1\}$ is necessary to decouple the
(implicit) optimization over $\bm{a}_m$ into the first $K-1$ individual
components $a_k$ and the magnitude $\norm{\bm{a}_m}^2$.

Define 
\begin{align*}
    \tilde{\bm{h}}_m & \defeq \frac{1}{\norm{\bm{h}_m}}
    \begin{pmatrix}
        h_{m,1} & \cdots & h_{m,K-1}
    \end{pmatrix}, \\
    q_m & \defeq \norm{\bm{a}_m}^2\in\N, \\
    \psi_{m}(q_m) & \defeq
    \max_{k\in\{1,\ldots, K-1\}} 
    \big\lvert\tilde{h}_{m,k}-a_{m,k}/\sqrt{q_m}\big\rvert.
\end{align*}
With this, we can rewrite \eqref{eq:thm_lattice1_0} as
\begin{equation}
    \label{eq:lattice1_1}
    \norm{\bm{a}_m}^2 + P\Bigl(\norm{\bm{h}_m}^2\norm{\bm{a}_m}^2 
    - \bigl( \bm{h}_m\bm{a}_m^{\T} \bigr)^2\Bigr)
    \geq q_m+\frac{4}{\pi^2}P\norm{\bm{h}_m}^2 
    q_m \psi_{m}^2(q_m).
\end{equation}
We want to minimize the rate loss \eqref{eq:lattice1_1}. The first term
in the right-hand side of \eqref{eq:lattice1_1} is increasing in $q_m$.
As we shall see, the second term is decreasing in $q_m$. Hence there is
a tradeoff between the two terms that determines the optimal value of
$q_m$. 

The behavior of the approximation error $\psi_m(q_m)$ can be bounded
using the convergent part of Khinchin's theorem in $K-1$ dimensions. To
this end, note that
\begin{align*}
    \psi_m(q_m)
    & \geq \min_{\bm{a}_m\in\Z^{K-1}}\max_{k\in\{1,\ldots,K-1\}}
    \big\lvert\tilde{h}_{m,k}-a_{m,k}/\sqrt{q_m}\big\rvert \\
    & = \max_{k\in\{1,\ldots,K-1\}} \min_{a_{m,k}\in\Z}
    \big\lvert\tilde{h}_{m,k}-a_{m,k}/\sqrt{q_m}\big\rvert,
\end{align*}
which is of the form analyzed in Lemma~\ref{thm:khinchin1}.  Applying
Lemma~\ref{thm:khinchin1} in $K-1$ dimensions shows then that for any
fixed $\delta > 0$ and almost every $\tilde{\bm{h}}_m$ there exists
$c=c(K,\tilde{\bm{h}}_m)=c(K,\bm{h})>0$ such that
\begin{equation}
    \label{eq:khinchine}
    \psi_m(q_m)
    > cq_m^{-1/2-1/(K-1)-\delta}
\end{equation}
for all $q_m \in\N$. Observe that this lower bound holds for any choice
of $\bm{a}_m$; in particular, the constant $c$ is uniform in $\bm{a}_m$.
We can then continue to lower bound the loss term in
$\RL(\bm{h}_m,P,\bm{a}_m)$ as
\begin{align}
    \label{eq:loss2}
    q_m + \frac{4}{\pi^2} P \norm{\bm{h}_m}^2
    q_m \psi_{m}^2(q_m)
    & \geq q_m + \frac{4}{\pi^2}c P \norm{\bm{h}_{m}}^2 q_m^{-2/(K-1)-2\delta} \nonumber\\
    & \geq \max\biggl\{q_m,\frac{4}{\pi^2}c P \norm{\bm{h}_{m}}^2 q_m^{-2/(K-1)-2\delta}\biggr\}. 
\end{align}  

This shows the tradeoff between the two cost terms. Recall that we are
allowed to choose $\bm{A}=\bm{A}(P)$, and hence also $q_m$, as a
function of power $P$. Asymptotically, the optimal choice of $q_m$ is
\begin{equation*}
    q_m = q_m(P) = \Theta\bigl(P^{(1+2/(K-1)+2\delta)^{-1}}\bigr),
\end{equation*}
and hence
\begin{equation*}
    q_m + P \norm{\bm{h}_m}^2 \frac{4}{\pi^2}q_m\psi_m^2(q_m)
    \geq \Omega\bigl(P^{(1+2/(K-1)+2\delta)^{-1}}\bigr)
\end{equation*}
as $P\to\infty$. 

Combined with \eqref{eq:lattice1_1}, this shows that, for almost every
$\tilde{\bm{h}}_m$, 
\begin{align*}
    \RL(\bm{h}_m, P, \bm{a}_m(P)) 
    & \leq \frac{1}{2}\log\bigl(1+P\norm{\bm{h}_m}^2\bigr)
    - \frac{1}{2}\log\bigl(\Omega\bigl(P^{(1+2/(K-1)+2\delta)^{-1}}\bigr)\bigr)
\end{align*}
as $P\to\infty$. This implies that
\begin{equation}
    \label{eq:lattice1_3b}
    \limsup_{P\to\infty} \frac{\RL(\bm{h}, P, \bm{a}_m(P))}{\frac{1}{2}\log(P)} 
    \leq \frac{2+\tilde{\delta}}{K+1+\tilde{\delta}},
\end{equation}
where we have set
\begin{equation*}
    \tilde{\delta} \defeq 2(K-1)\delta > 0.
\end{equation*}
We have argued that \eqref{eq:lattice1_3b} holds for almost every
$\tilde{\bm{h}}_m\in\R^{K-1}$. It is shown in Appendix~\ref{sec:measure}
that this implies that \eqref{eq:lattice1_3b} also holds for almost
every $\bm{h}_m\in\R^K$. 

Up to this point, we have analyzed the rate for a single receiver $m$.
Using the definition of $\RL(\bm{H}, P, \bm{A})$ in
\eqref{eq:rl4}, this yields the upper bound
\begin{align*}
    \RL(\bm{H}, P, \bm{A}(P))
    & = \sum_{k=1}^K \min_{m: a_{m,k}\neq 0} 
    \RL(\bm{h}_{m}, P, \bm{a}_{m}(P)) \\
    & \leq K \max_{m\in\{1,\ldots, K\}} \RL(\bm{h}_m, P, \bm{a}_m(P))
\end{align*}
on the sum rate. Together with \eqref{eq:lattice1_3b} and using the
union bound over $m\in\{1,\ldots, K$\}, this implies that
\begin{equation*}
    \limsup_{P\to\infty} \frac{\RL(\bm{H}, P, \bm{A}(P))}
    {\frac{1}{2}\log(P)} \leq \frac{2+\tilde{\delta}}{1+1/K+\tilde{\delta}/K}
\end{equation*}
for almost every $\bm{H}$. As we have assumed that $\bm{A}(P)$ is the
optimal coefficient matrix for power $P$, this implies by \eqref{eq:rl3}
that
\begin{equation*}
    \limsup_{P\to\infty} \frac{\RL(\bm{H}, P)}
    {\frac{1}{2}\log(P)} \leq \frac{2+\tilde{\delta}}{1+1/K+\tilde{\delta}/K}
\end{equation*}
for almost every $\bm{H}$. Since $\tilde{\delta}> 0$ is arbitrary, we
may take the limit as $\tilde{\delta}\to 0$ to obtain 
\begin{equation*}
    \limsup_{P\to\infty} \frac{\RL(\bm{H}, P)}
    {\frac{1}{2}\log(P)} \leq \frac{2}{1+1/K},
\end{equation*}
yielding the desired upper bound on the degrees of freedom of the
lattice scheme.

We now derive an estimate of the speed of convergence. Observe that we can choose 
$\delta$ in \eqref{eq:khinchine} as
\begin{equation*}
    \delta 
    = \delta(q_m) 
    = \frac{2\log\log(1+q_m)}{(K-1)\log(q_m)} 
\end{equation*}
and still satisfy the convergence condition in Lemma~\ref{thm:khinchin1}
in $K-1$ dimensions. The lower bound \eqref{eq:loss2} on the loss term
in $\RL(\bm{h}_m,P,\bm{a}_m)$ then becomes
\begin{align*}
    q_m + \frac{4}{\pi^2}P\norm{\bm{h}_m}^2 
    q_m \psi_{m}^2(q_m)
    & \geq \max\biggl\{q_m,
    \frac{4}{\pi^2}c P \norm{\bm{h}_{m}}^2 q_m^{-\tfrac{2}{K-1}}\log^{-\tfrac{4}{K-1}}(1+q_m)\biggr\} \\
    & \geq \log^{-\tfrac{4}{K-1}}(1+q_m)\max\biggl\{q_m,
    \frac{4}{\pi^2}c P \norm{\bm{h}_{m}}^2 q_m^{-\tfrac{2}{K-1}}\biggr\}.
\end{align*}  
By \cite[Lemma~1]{nazer11}, we can restrict the optimization over
$\bm{A}$ to matrices satisfying 
\begin{equation*}
    q_m = \norm{\bm{a}_m}^2 \leq \norm{\bm{h}_m}^2P
\end{equation*}
so that
\begin{align*}
    q_m + \frac{4}{\pi^2}P \norm{\bm{h}_m}^2 
    q_m \psi_{m}^2(q_m)
    & \geq \log^{-\tfrac{4}{K-1}}(1+\norm{\bm{h}_m}^2 P) 
    \max\biggl\{q_m, \frac{4}{\pi^2}cP \norm{\bm{h}_{m}}^2 
    q_m^{-\tfrac{2}{K-1}}\biggr\}. 
\end{align*}  
We can now solve for the optimal $q_m$. Proceeding as before, we obtain
an upper bound on the computation rate with lattice coding of
\begin{align*}
    \RL(\bm{H}, P) 
    & \leq \max_{m\in\{1,\ldots, K\}}K
    \Biggl(
    \frac{1}{2}\log\bigl(1+P\norm{\bm{h}_m}^2\bigr) 
    - \frac{1}{2}\log
    \Omega\biggl(
    \log^{-\tfrac{4}{K-1}}(1+\norm{\bm{h}_m}^2P)P^{\bigl(1+\tfrac{2}{K-1}\bigr)^{-1}}
    \biggr) 
    \Biggr) \nonumber\\
    & \leq \frac{1}{1+1/K}\log(P)
    + O\bigl(\log\log(P)\bigr)
\end{align*}
as $P\to\infty$. This implies that for any $K\geq 2$ and
almost every $\bm{H}\in\R^{K\times K}$ there exists a positive constant
$c_1 = c_1(K,\bm{H})$ such that for all $P \geq 3$
\begin{equation*}
    \RL(\bm{H}, P)
    \leq \frac{1}{1+1/K}\log(P) + c_1\log\log(P),
\end{equation*}
proving the theorem.  \hfill\IEEEQED

\section{Proof of Theorem~\ref{thm:lattice3}}
\label{sec:proofslattice3}

The upper bound in Theorem~\ref{thm:lattice3} follows immediately from
\eqref{eq:upper}. We focus here on the lower bound showing that
\begin{equation*}
    \liminf_{P\to\infty}\frac{C(\bm{H},P)}{\frac{1}{2}\log(P)} \geq K.
\end{equation*}
We start with a high-level description of the scheme achieving this
performance in Section~\ref{sec:proofslattice3_scheme}. The detailed
analysis can be found in Section~\ref{sec:proofslattice3_details}.

\subsection{Description of Communication Scheme}
\label{sec:proofslattice3_scheme}

\begin{figure*}[htbp]
    \begin{center}
        \hspace{-0.4cm}\scalebox{0.667}{\input{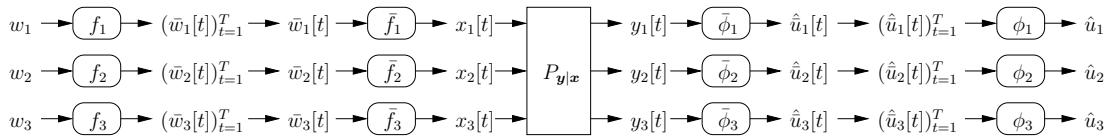}} 
    \end{center}

    \caption{Modulation scheme $(\{\bar{f}_k\}, \{\bar{\phi}_m\})$
    together with outer code $(\{f_k\}, \{\phi_m\})$. At transmitter
    $k$, the message $w_k$ is mapped by the encoder $f_k$ of the outer
    code into the sequence $(\bar{w}_k[t])_{t=1}^T$ of modulator inputs.
    Each $\bar{w}_k[t]$ is mapped by the modulator $\bar{f}_k$ to a
    channel input $x_k[t]$. At receiver $m$, the channel output $y_m[t]$
    is mapped by the demodulator $\bar{\phi}_m$ into a demodulated
    equation $\hat{\bar{u}}_m[t]$. The sequence of these demodulated
    equations $(\hat{\bar{u}}_m[t])_{t=1}^T$ is mapped by the decoder
    $\phi_m$ of the outer code to an estimate $\hat{u}_m$ of the desired
    equation $u_m$.}

    \label{fig:scheme}
\end{figure*}

The proposed coding scheme consists of two components: a modulation
scheme and an outer code (see Fig.~\ref{fig:scheme}). The encoder $f_k$
for the outer code at transmitter $k$ maps the message $w_k$ into the
sequence of coded symbols $(\bar{w}_k[t])_{t=1}^T$. The modulator
$\bar{f}_k$ at transmitter $k$ maps each coded symbol $\bar{w}_k[t]$
into a channel symbol $x_k[t]$.  Thus, while the outer code produces a
block of coded symbols, the modulation scheme operates on a single coded
symbol to produce a single channel symbol. The encoder in the definition
of computation capacity is the concatenation of these two encoding
operations. At receiver $m$, the demodulator $\bar{\phi}_m$ computes
$\hat{\bar{u}}_m[t]$ from the channel output $y_m[t]$, and the decoder
$\phi_m$ for the outer code maps the sequence
$(\hat{\bar{u}}_m[t])_{t=1}^T$ into an estimate $\hat{u}_m$ of the
desired function $u_m$. Both $u_m$ and $\bar{u}_m$ are defined as a
function of $(w_k)$ and $(\bar{w}_k)$, respectively.  The decoder in the
definition of computation capacity is the concatenation of these two
decoding operations.  

Note that in the description of the proposed achievable scheme we are
using the following notational conventions. Quantities related to the
outer code are denoted by standard font, i.e., $f_k$, $\phi_m$, $w_k$,
\ldots The corresponding quantities related to the modulation scheme are
indicated by bars, i.e., $\bar{f}_k$, $\bar{\phi}_m$, $\bar{w}_k$,
\ldots Estimated quantities are indicated by hats, i.e., the output
$\hat{u}_m$ of the decoder of the outer code is an estimate of the
correct output $u_m$, and similarly the output $\hat{\bar{u}}_m$ of the
demodulator is an estimate of the correct output $\bar{u}_m$.

The construction is as follows. Each message $w_k$ is split into
$\card{\mc{G}_L}=L^{K^2}$ submessages $(w_{k,g})_{g\in\mc{G}_L}$ with
$\mc{G}_L$ defined in \eqref{eq:mcgl}. Every $f_k$ encodes each of
these submessages $w_{k,g}$ using the same linear code. Thus, all
encoders $\{f_k\}$ are identical. The modulator $\bar{f}_k$ combines
these $\card{\mc{G}_L}$ codewords into a single sequence of channel
inputs. 

Consider now receiver $m$. The channel to this receiver is in effect a
$K$-user multiple-access channel (MAC). By splitting the transmitted
message into submessage, we have transformed this $K$-user MAC into a
$K\card{\mc{G}_L}$-user MAC, with each user corresponding to one
submessage. The demodulator $\bar{\phi}_m$ splits this MAC at receiver
$m$ into $\card{\mc{G}_{L+1}}$ subchannels. Through careful design of
the modulators, this splitting can be done such that each of the
resulting $\card{\mc{G}_{L+1}}$ MACs outputs a (noisy) sum of only $K$
out of the $K\card{\mc{G}_L}$ possible input signals. Observe that this
channel is linear with integer channel coefficients. Hence, the linear
codes used as outer code can now be efficiently decoded. The decoder
$\phi_m$ of the outer code is thus chosen to recover the submessage
corresponding to the \emph{sum} of the $K$ codewords seen over this MAC.
Decoding is shown to be possible with vanishing probability of error
with a rate of order $\tfrac{1}{2\card{\mc{G}_{L+1}}}\log(P)$ for each
of the submessages for large $P$.  Moreover, it can be shown that the
resulting collection of decoded functions is invertible.

Since there are $\card{\mc{G}_L}$ submessages for each of the $K$
transmitters, the sum rate achieved by this scheme is on the order of
\begin{equation*}
    \frac{K\card{\mc{G}_L}}{2\card{\mc{G}_{L+1}}}\log(P)
    = \frac{K}{2(1+1/L)^{K^2}}\log(P).
\end{equation*}
The scheme achieves therefore
\begin{equation*}
    \frac{K}{(1+1/L)^{K^2}}
\end{equation*}
degrees of freedom. For large $L$, this is approaches the $K$ degrees of
freedom claimed in Theorem~\ref{thm:lattice3}.

\subsection{Detailed Proof of Achievability}
\label{sec:proofslattice3_details}

The proof of the theorem consists of three steps. First, we show how the
modulation scheme transforms the \emph{noisy} linear combinations with
\emph{real} coefficients produced by the channel \eqref{eq:channel} into
a system computing noisy linear combinations with \emph{integer}
coefficients. Second, we show how the outer code further transforms this
modulated channel into a system computing \emph{noiseless} linear
combinations with integer coefficients. Third, we argue that the linear
combinations produced by the outer code are invertible, i.e., the
messages at the transmitters can be recovered from the computed linear
combinations of all receivers.

\begin{figure}[htbp]
    \begin{center}
        \hspace{-1.5cm}\scalebox{0.667}{\input{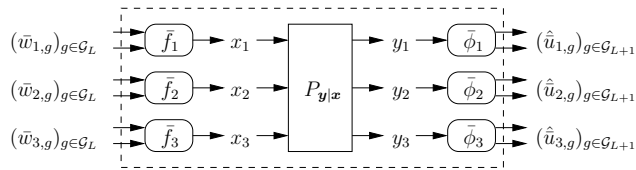}} 
    \end{center}

    \caption{Modulation scheme $(\{\bar{f}_k\}, \{\bar{\phi}_m\})$.  The
    modulator $\bar{f}_k$ at transmitter $k$ takes
    $(\bar{w}_{k,g})_{g\in\mc{G}_L}$ as input. The demodulator
    $\bar{\phi}_m$ at receiver $m$ produces
    $(\hat{\bar{u}}_{m,g})_{g\in\mc{G}_{L+1}}$ as its output.
    Indicated by the dashed box is the modulated channel obtained by
    viewing the modulation scheme as part of the channel. This modulated
    channel is discrete and memoryless. All operations take place over a
    single time slot $t$; the dependence of $\bar{w}_{k,g}$, $x_k$,
    $y_m$, $\hat{\bar{u}}_{m,g}$ on $t$ is omitted in the figure.}

    \label{fig:modulation}
\end{figure}

We now describe the operations of the modulation scheme in detail (see
Fig.~\ref{fig:modulation}).  Recall the definition of $\mc{G}_L$ in
\eqref{eq:mcgl} as the collection of all monomials in the channel gains
with exponents between $0$ and $L-1$. The input symbol $\bar{w}_k[t]$ to
the modulator $\bar{f}_k$ at transmitter $k$ at time $t$ consists of
$\card{\mc{G}_L}$ subsymbols 
\begin{equation*}
    \bar{w}_k[t] 
    \defeq (\bar{w}_{k,g}[t])_{g\in\mc{G}_L}, 
    \quad \bar{w}_{k,g}[t]\in\{0,\ldots, p-1\} \ \forall k,g,t
\end{equation*}
for some $p,L\in\N$ to be chosen later.
The output symbol $\hat{\bar{u}}_m[t]$ of the demodulator $\bar{\phi}_m$ at
receiver $m$ at time $t$ consists of $\card{\mc{G}_{L+1}}$ subsymbols 
\begin{equation*}
    \hat{\bar{u}}_m[t] 
    \defeq (\hat{\bar{u}}_{m,g}[t])_{g\in\mc{G}_{L+1}},
    \quad \hat{\bar{u}}_{m,g}[t]\in\{0,\ldots, p-1\} \ \forall m,g,t.
\end{equation*}
Note that the number of input and output subsymbols per time slot are
not the same.

The modulator $\bar{f}_k$ at transmitter $k$ is a linear map, producing
the channel input
\begin{equation}
    \label{eq:xdef}
    x_k[t] 
    \defeq \bar{f}_k\bigl( (\bar{w}_{k,g}[t])_{g\in\mc{G}_L} \bigr)
    \defeq B \sum_{g\in\mc{G}_L} \bar{w}_{k,g}[t] g
\end{equation}
with 
\begin{equation}
    \label{eq:bdef}
    B 
    = B(L,p) 
    \defeq (Kp)^{\card{\mc{G}_{L+1}}}.
\end{equation}
For $g\in\mc{G}_{L+1}$, define
\begin{equation}
    \label{eq:ubardef}
    \bar{u}_{m,g}[t] 
    \defeq \sum_{k=1}^K \bar{w}_{k,(g/h_{m,k})}[t],
\end{equation}
where we use the convention that $\bar{w}_{k,(g/h_{m,k})}[t]=0$ whenever
$g/h_{m,k}\notin\mc{G}_{L}$. 

The definition of $\bar{u}_{m,g}[t]$ can be interpreted in the following
way. Let $\tilde{g}\in\mc{G}_L$, and consider the term
$\bar{w}_{k,\tilde{g}}[t]\tilde{g}$ in the definition of $x_k[t]$. At
receiver $m$, this term is observed as $\bar{w}_{k,\tilde{g}}[t]
\tilde{g} h_{m,k}$. Thus, for any $g\in\mc{G}_{L+1}$, $\bar{u}_{m,g}[t]$
is the sum of all input subsymbols
$(\bar{w}_{k,\tilde{g}}[t])_{\tilde{g}\in\mc{G}_L}$ that are observed
with coefficient $g$ at receiver $m$. Another way to see this is as
follows. The signal observed at receiver $m$ is
\begin{equation*}
    y_m 
    = B\sum_{k=1}^K \sum_{g\in\mc{G}_L} \bar{w}_{k,g}[t] h_{m,k}g +z_m.
\end{equation*}
Now, note that $h_{m,k}g$ is a monomial in the channel gains with
highest exponent at most $L$. Hence $h_{m,k}g\in\mc{G}_{L+1}$ for all
$m,k$. Using the definition of $\bar{u}_{m,g}$, we can rewrite the
received signal as
\begin{align*}
    y_m 
    & = B \sum_{k=1}^K \sum_{g\in\mc{G}_{L+1}} \bar{w}_{k,(g/h_{m,k})}[t] g +z_m \\
    & = B \sum_{g\in\mc{G}_{L+1}}\bar{u}_{m,g}[t]g +z_m.
\end{align*}
This last equation is a key step in the construction of the achievable
scheme. It shows that the received signal can be decomposed into
$\card{\mc{G}_{L+1}}$ terms $\bar{u}_{m,g}$, each multiplied by a
different effective channel gain $g$. Crucially, each of these terms is
an \emph{integer} linear combination of up to $K$ input signals
$\bar{w}_{k,(g/h_{m,k})}$, one from each transmitter.

The demodulator $\bar{\phi}_m$ at receiver $m$ is the maximum likelihood
detector of $(\bar{u}_{m,g}[t])_{g\in\mc{G}_{L+1}}$, i.e., 
\begin{equation*}
    \bar{\phi}_m(y_m[t])
    \defeq \argmax_{(\hat{\bar{u}}_{m,g})} 
    \Pp\bigl(\cap_{g\in\mc{G}_{L+1}} \{\bar{u}_{m,g}[t] = \hat{\bar{u}}_{m,g}\}
    \bigm\vert y_m[t] \bigr),
\end{equation*}
where the arg max is over all possible values of
$(\hat{\bar{u}}_{m,g})_{g\in\mc{G}_{L+1}}$. Denote by
\begin{equation*}
    (\hat{\bar{u}}_{m,g}[t])_{g\in\mc{G}_{L+1}} 
    \defeq \bar{\phi}_m(y_m[t])
\end{equation*} 
the output of the demodulator. The \emph{probability of demodulation
error} at receiver $m$ is then defined as
\begin{align*}
    \Pp\bigl( \cup_{g\in\mc{G}_{L+1}} 
    \{ \hat{\bar{u}}_{m,g}[t] \neq \bar{u}_{m,g}[t]\}
    \bigr).
\end{align*}
Observe that the goal of the demodulator is to recover
$\card{\mc{G}_{L+1}}$ integers $(\bar{u}_{m,g}[t])_g$
from a single observation $y_m[t]$. 

The next lemma describes the performance of this modulation scheme.
\begin{lemma}
    \label{thm:modulation}
    For any $K\geq 2$ and almost every $\bm{H}\in\R^{K\times K}$ there
    exist positive constants $c_4=c_4(K,\bm{H})$ and $c_5=c_5(K,\bm{H})$
    such that, for all $p,L,t\in\N$ and $k\in\{1,\ldots, K\}$, the input
    signal to the channel has power at most
    \begin{align}
        \label{eq:power}
        x_k^2[t] 
        & \leq c_4^L (Kp)^{2\card{\mc{G}_{L+1}}} L^{2K^2}p^2 \nonumber\\
        & \defeq P(L,p)
    \end{align}
    and 
    \begin{equation*}
        \bigcup_{g\in\mc{G}_{L+1}} 
        \{ \hat{\bar{u}}_{m,g}[t]\neq \bar{u}_{m,g}[t] \}
        \subset
        \{\abs{z_m[t]} > c_5p\},
    \end{equation*}
    implying that the probability of demodulation error is at
    most
    \begin{align}
        \label{eq:eps}
        \Pp\bigl( \cup_{g\in\mc{G}_{L+1}} 
        \{ \hat{\bar{u}}_{m,g}[t]\neq \bar{u}_{m,g}[t] \} \bigr) 
        & \leq \Pp\bigl( \{\abs{z_m[t]} > c_5p^{1/2}\} \bigr) \nonumber\\
        & \leq \exp\bigl(- \tfrac{1}{2}c_5^2p\bigr) \nonumber\\
        & \defeq \varepsilon(p).
    \end{align}
\end{lemma}
The proof of Lemma~\ref{thm:modulation} is presented in
Section~\ref{sec:proofslattice3_modulation}.

Lemma~\ref{thm:modulation} bounds the power of the channel input
$x_k[t]$ and, more importantly, states that the probability of
demodulating in error decreases exponentially in $p$.  Thus, when
$p$ is large enough, the probability of demodulation error is small. 

The original channel between transmitters and receivers produces
\emph{noisy} \emph{real} linear combinations of the channel inputs.
After applying the modulation scheme, we have transformed this into a
channel that produces \emph{noisy} \emph{integer} linear combinations of
the channel inputs. More precisely, using the definition of
$\bar{u}_{m,g}$ in \eqref{eq:ubardef}, we can write this new channel as
\begin{align}
    \label{eq:modulated_channel}
    \hat{\bar{u}}_{m,g}[t]
    & = \bar{u}_{m,g}[t] + \bar{z}_{m,g}[t] \nonumber\\
    & = \sum_{k=1}^K \bar{w}_{k,(g/h_{m,k})}[t] + \bar{z}_{m,g}[t],
\end{align}
where we have defined the \emph{modulation noise} $\bar{z}_{m,g}[t]$ as
\begin{equation}
    \label{eq:zmgdef}
    \bar{z}_{m,g}[t] \defeq \hat{\bar{u}}_{m,g}[t]- \bar{u}_{m,g}[t].
\end{equation}
Thus, we see that the channel resulting between the input of the
modulator and the output of the demodulator computes noisy linear
combinations with integer (indeed, either zero or one)
coefficients. 

We refer to this new channel after modulation as the \emph{modulated
channel}. Since the modulation scheme operates on a single time slot
$t$, this modulated channel is discrete and memoryless. Note that the
noise $\bar{z}_{m,g}$ of this modulated channel is not necessarily
additive, i.e., $\bar{z}_{m,g}[t]$ is not necessarily independent of the
channel input $\bar{u}_{m,g}[t]$. However, by
Lemma~\ref{thm:modulation}, we know that the noise is small.  This
modulated channel is depicted in Fig.~\ref{fig:modulation}.

As we have argued before, the probability of demodulation error
$\varepsilon(p)$ goes to zero as $p\to\infty$ and hence, by
\eqref{eq:power}, as power $P\to\infty$.  However, the definition of
computation capacity requires that the probability of error be
arbitrarily small for \emph{fixed} power $P$.  The next step is
therefore to transform the modulated channel into a system producing
\emph{noiseless} \emph{integer} linear combinations of the channel
inputs. To this end, we employ an outer code over the modulated channel.
We call the encoder and decoder of this channel code $f_k$ and $\phi_m$
for transmitter $k$ and receiver $m$, respectively. It will be
convenient to choose $p$ to be a prime number. Reducing the modulator
outputs modulo $p$, we can then interpret the input and output
subsymbols $\bar{w}_{k,g}[t]$ and $\hat{\bar{u}}_{m,g}[t]$ as well as
the integer linear combinations performed by the channel as being in the
finite field $\F_p$. We refer to the resulting channel over $\F_p$ as
the \emph{modulated $\F_p$ channel}.

\begin{figure*}[htbp]
    \begin{center}
        \hspace{-1.5cm}\scalebox{0.667}{\input{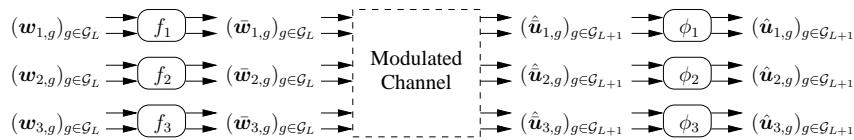}} 
    \end{center}

    \caption{Outer code $(\{f_k\}, \{\phi_m\})$ over the modulated
    $\F_p$ channel. Each encoder $f_k$ uses the same linear map given by
    the matrix $\bm{S}\in\F_p^{T\times TR/\log(p)}$. $\phi_m$ is the
    corresponding minimum distance decoder for the equations
    $\bm{u}_m$.}

    \label{fig:outer}
\end{figure*}

We now describe the operations of the outer code in more detail (see
Fig.~\ref{fig:outer}). The channel encoder $f_k$ at transmitter $k$
consists of $\card{\mc{G}_L}$ sub-encoders 
\begin{equation*}
    f_k \defeq \{f_{k,g}\}_{g\in\mc{G}_L}.
\end{equation*}
The channel decoder $g_m$ at receiver $m$ consists of
$\card{\mc{G}_{L+1}}$ sub-decoders
\begin{equation*}
    \phi_m \defeq \{\phi_{m,g}\}_{g\in\mc{G}_{L+1}}.
\end{equation*}

Consider the message $w_{k,g}$ at sub-encoder $g$ of transmitter $k$.
It will be convenient in the following to express this message as a
vector in $\F_p^{TR/\log(p)}$. Whenever this vector structure is
relevant, we will write the message as $\bm{w}_{k,g}$. The encoder
$f_{k,g}$ maps $\bm{w}_{k,g}$ to the vector (or, equivalently, sequence)
of modulator inputs 
\begin{equation*}
    \bar{\bm{w}}_{k,g}
    \defeq (\bar{w}_{k,g}[t])_{t=1}^T
    \in\F_p^T.
\end{equation*}
The rate of this channel encoder is hence
\begin{equation*}
    \frac{\log(p^{TR/\log(p)})}{T} = R
\end{equation*}
bits per use of the modulated channel. The encoder is specified by the
linear map 
\begin{equation}
    \label{eq:fkgdef}
    \bar{\bm{w}}_{k,g}
    \defeq f_{k,g}(\bm{w}_{k,g})
    \defeq \bm{S}\bm{w}_{k,g},
\end{equation}
for some matrix $\bm{S}\in\F_p^{T\times TR/\log(p)}$, and where the
multiplication is understood to be over $\F_p$.  We point out that
$\bm{S}$ does not depend on $k$ and $g$. In other words, each encoder
$f_{k,g}$ uses the \emph{same} linear map.

Define the vector version of the demodulator output $\hat{\bar{u}}_{m,g}$
as
\begin{equation*}
    \hat{\bar{\bm{u}}}_{m,g} 
    \defeq (\hat{\bar{u}}_{m,g}[t])_{t=1}^T \in\F_p^T.
\end{equation*}
Similar to the definition of $\bar{u}_{m,g}[t]$ in \eqref{eq:ubardef},
set
\begin{equation}
    \label{eq:bmu}
    \bm{u}_{m,g}
    \defeq \sum_{k=1}^K \bm{w}_{k,(g/h_{m,k})} \pmod{p},
\end{equation}
where we again use the convention that $\bm{w}_{m,(g/h_{m,k})}=\bm{0}$
whenever $g/h_{m,k}\notin\mc{G}_{L}$. 

Recall that the modulated channel computes noisy linear combinations
over the finite field $\F_p$. Since all channel encoders use the same
linear code, this implies that the output of the subchannel $g$ at
receiver $m$ is equal to $\bm{S}\bm{u}_{m,g}$ plus small noise
$\bar{\bm{z}}_{m,g}$ resulting from erroneous demodulation as defined in
\eqref{eq:zmgdef}. As pointed out earlier, the noise term
$\bar{\bm{z}}_{m,g}$ may not be additive, i.e., $\bar{\bm{z}}_{m,g}$ may
be dependent on the channel inputs. Formally, the (vector) of
demodulated equations $\hat{\bar{\bm{u}}}_{m,g}$ is equal to
\begin{align}
    \label{eq:linear}
    \hat{\bar{\bm{u}}}_{m,g}
    & \stackrel{(a)}{=} \sum_{k=1}^K \bar{\bm{w}}_{k,(g/h_{m,k})}+\bar{\bm{z}}_{m,g} \nonumber\\
    & \stackrel{(b)}{=} \sum_{k=1}^K \bm{S}\bm{w}_{k,(g/h_{m,k})}+\bar{\bm{z}}_{m,g} \nonumber\\
    & = \bm{S}\sum_{k=1}^K \bm{w}_{k,(g/h_{m,k})}+\bar{\bm{z}}_{m,g} \nonumber\\
    & \stackrel{(c)}{=} \bm{S}\bm{u}_{m,g}+\bar{\bm{z}}_{m,g} \pmod{p},
\end{align}
where $(a)$ follows \eqref{eq:modulated_channel}, $(b)$ follows from the
definition of the encoder $f_{k,g}$ in \eqref{eq:fkgdef}, and $(c)$
follows from the definition of the equation $\bm{u}_{m,g}$ in
\eqref{eq:bmu}. Thus, since all transmitters use the \emph{same linear}
code, the encoding operation commutes with the operation of the channel.
Note that \eqref{eq:ubardef} and \eqref{eq:linear} imply that
\begin{equation}
    \label{eq:barbmu2}
    \bar{\bm{u}}_{m,g} = \bm{S}\bm{u}_{m,g}.
\end{equation}

The decoder $\phi_{m,g}$ of the outer code is the minimum (Hamming)
distance decoder, i.e., 
\begin{equation*}
    \phi_{m,g}(\hat{\bar{\bm{u}}}_{m,g})
    \defeq \argmin_{\hat{\bm{u}}_{m,g}\in\F_p^{TR/\log(p)}} 
    \sum_{t=1}^T
    \ind\bigl(\hat{\bar{u}}_{m,g}[t] \neq (\bm{S}\hat{\bm{u}}_{m,g})[t]\bigr),
\end{equation*}
where $(\bm{S}\hat{\bm{u}}_{m,g})[t]$ is component $t$ of the vector $
\bm{S}\hat{\bm{u}}_{m,g}$. Note that this decoder might not be the same
as the maximum likelihood decoder, depending on the distribution of
$\bar{\bm{z}}_{m,g}$. Denote by
\begin{equation*}
    \hat{\bm{u}}_{m,g} \defeq
    \phi_{m,g}(\hat{\bar{\bm{u}}}_{m,g})
\end{equation*}
the output of the decoder of the outer code.
The \emph{probability of error} of this code is defined as
\begin{equation*}
    \Pp\bigl(\cup_{m,g} \{\hat{\bm{u}}_{m,g} \neq \bm{u}_{m,g}\} \bigr).
\end{equation*}

For $x\in(0,1)$, define the $p$-ary entropy function $\mc{H}_p(x)$ as
\begin{equation}
    \label{eq:entropy}
    \mc{H}_p(x) 
    \defeq \frac{1}{\log(p)}\bigl(x\log(p-1)-x\log(x)-(1-x)\log(1-x)\bigr).
\end{equation}
The next lemma states that for the modulated $\F_p$ channel there exist
linear codes $\bm{S}$ with large rate that allow reliable decoding at
each receiver.

\begin{lemma}
    \label{thm:coding}
    Denote by $\varepsilon(p)$ the upper bound on the probability of
    demodulation error as defined in \eqref{eq:eps} in
    Lemma~\ref{thm:modulation}. For every prime number $p$ such that
    $\varepsilon(p) < 1/4$, and every $\eta\in(0, 1/2-2\varepsilon(p))$
    there exists a linear code $\bm{S}$ (of sufficiently large
    blocklength $T$) for the modulated $\F_p$ channel with rate bigger
    than
    \begin{equation*}
        \bigl(1-\mc{H}_p(2\varepsilon(p)+\eta)\bigr)\log(p)
    \end{equation*}
    and probability of error less than $\eta$. 
\end{lemma}
The proof of Lemma~\ref{thm:coding} is presented in
Section~\ref{sec:proofslattice3_coding}.

Lemma~\ref{thm:coding} shows that, asymptotically in the blocklength
$T$, reliable communication over the modulated subchannels is possible
at rates arbitrarily close to 
\begin{equation*}
    \bigl(1-\mc{H}_p(2\varepsilon(p))\bigr)\log(p),
\end{equation*}
with probability of demodulation error $\varepsilon(p)$ as defined in
Lemma~\ref{thm:modulation}. Since there are $K$ transmitters each with
$\card{\mc{G}_L}$ subchannels, the sum rate achieved with this coding
scheme is at least
\begin{equation*}
    K\card{\mc{G}_{L}}\bigl(1-\mc{H}_p(2\varepsilon(p))\bigr)\log(p).
\end{equation*}
Note that
\begin{equation*}
    \lim_{p\to\infty}\mc{H}_p(2\varepsilon(p)) = 0
\end{equation*}
so that the sum rate is of order
\begin{equation*}
    K\card{\mc{G}_{L}}(1-o(1))\log(p)
\end{equation*}
as $p\to\infty$.

To satisfy the definition of computation capacity, we need to argue that
the mapping from $(\bm{w}_{k,g})$ to $(\bm{u}_{m,g})$ defined in
\eqref{eq:bmu} is deterministic and invertible over its range. As the
channel gains $\bm{H}$ are constant and known, the mapping is clearly
deterministic.  The next lemma shows that the mapping is also injective
(and hence invertible over its range).

\begin{lemma}
    \label{thm:invertible}
    Let $p$ be a prime number.  For any $K\geq 2$ and almost every
    $\bm{H}\in\R^{K\times K}$, the mapping from 
    \begin{equation*}
        \bigl(\bm{w}_{k,g}: k\in\{1,\ldots, K\}, g\in\mc{G}_L\bigr)
    \end{equation*}
    to
    \begin{equation*}
        \bigl(\bm{u}_{m,g}: m\in\{1,\ldots, K\}, g\in\mc{G}_{L+1}\bigr)
    \end{equation*}
    is injective over $\F_p$.
\end{lemma}
The proof of Lemma~\ref{thm:invertible} is presented in
Section~\ref{sec:proofslattice3_invertible}.

Together with Lemmas~\ref{thm:modulation} and \ref{thm:coding},
Lemma~\ref{thm:invertible} shows that for every prime number $p$, a
computation rate
\begin{equation}
    \label{eq:achievable}
    C(\bm{H},P(L,p)) \geq K\card{\mc{G}_{L}}
    \bigl(1-\mc{H}_p(2\varepsilon(p))\bigr)\log(p)
\end{equation}
is achievable, with 
\begin{equation*}
    P(L,p) = c_4^L (Kp)^{2\card{\mc{G}_{L+1}}}L^{2K^2}p^2
\end{equation*}
as defined in \eqref{eq:power}. 

Fix a power $P$, and let $p$ and $\tilde{p}$ be two consecutive prime
numbers such that
\begin{equation*}
    P(L,p) \leq P \leq P(L,\tilde{p}).
\end{equation*} 
By Bertrand's postulate (see, for example, \cite[Theorem~5.7.1]{hua82}),
any two consecutive primes $p$ and $\tilde{p}$ satisfy, 
\begin{equation*}
    p < \tilde{p} \leq 2 p.
\end{equation*}
Since $P(L,p)$ is increasing in $p$, this implies that
\begin{equation}
    \label{eq:power2}
    P(L,p) \leq P \leq P(L,2p).
\end{equation}

Combining \eqref{eq:achievable} and \eqref{eq:power2} shows that, for
every power $P$ and corresponding prime number $p$ chosen as above,
\begin{align*}
    \frac{C(\bm{H},P)}{\frac{1}{2}\log(P)} 
    \geq \frac{C(\bm{H},P(L,p))}{\frac{1}{2}\log(P(L,2p))}
    = D(p),
\end{align*}
where
\begin{align*}
    D(p)
    & \defeq \frac{K\card{\mc{G}_{L}}\bigl(1-\mc{H}_p(2\varepsilon(p))\bigr)\log(p)}
    {\frac{1}{2}L\log(c_4)+\card{\mc{G}_{L+1}}\log(2Kp)
    + K^2\log(L)+\log(2p)}.
\end{align*}
Since $p\to\infty$ as $P\to\infty$ (with $K$ and $L$ fixed), this
implies that
\begin{align*}
    \liminf_{P\to\infty}\frac{C(\bm{H},P)}{\frac{1}{2}\log(P)} 
    \geq \lim_{p\to\infty} D(p) 
    = \frac{K\card{\mc{G}_{L}}}{\card{\mc{G}_{L+1}}+1},
\end{align*}
where the limit $p\to\infty$ is understood as being taken over the prime
numbers. By Remark~\ref{rem:unique} in
Section~\ref{sec:proofspreliminaries}, we have
\begin{equation*}
    \card{\mc{G}_L} = L^{K^2},
\end{equation*}
for almost every $\bm{H}$. Hence, this shows that
\begin{equation*}
    \liminf_{P\to\infty}\frac{C(\bm{H},P)}{\frac{1}{2}\log(P)} 
    \geq \frac{KL^{K^2}}{(L+1)^{K^2}+1}.
\end{equation*}
As this is true for all values of $L$, we may take the limit
$L\to\infty$ to obtain 
\begin{equation*}
    \liminf_{P\to\infty}\frac{C(\bm{H},P)}{\frac{1}{2}\log(P)} 
    \geq \lim_{L\to\infty}\frac{KL^{K^2}}{(L+1)^{K^2}+1}
    = K.
\end{equation*}
Hence, the proposed implementation of compute-and-forward achieves
$K$ degrees of freedom, as needed to be shown.

We now derive an estimate of the rate of convergence to this limiting value. 
Fix a power $\tilde{P}$. Set\footnote{The number $L(\tilde{P})$ might
not be an integer. We ignore the rounding error since it is 
immaterial as $\tilde{P}\to\infty$.}
\begin{equation*}
    L(\tilde{P}) \defeq \log^{\tfrac{1}{1+K^2}}(\tilde{P}),
\end{equation*}
and let $p(\tilde{P})$ be the largest prime number $p$ such that
$P(L(\tilde{P}), p) \leq \tilde{P}$. Using Bertrand's postulate as before, we obtain that
\begin{equation}
    \label{eq:Pbound}
    P(L(\tilde{P}), p(\tilde{P})) 
    \leq \tilde{P}
    \leq P(L(\tilde{P}), 2p(\tilde{P})).
\end{equation}

Solving $P(L,p)$ in \eqref{eq:power} for $p$ yields
\begin{equation*}
    p 
    = \biggl(
    \frac{P}{c_4^L K^{2\card{\mc{G}_{L+1}}}L^{2K^2}}
    \biggr)^{\tfrac{1}{2}(\card{\mc{G}_{L+1}}+1)^{-1}}.
\end{equation*}
Together with \eqref{eq:Pbound}, this implies that
\begin{equation}
    \label{eq:pbound}
    \log(p(\tilde{P}))
    = \frac{\log(\tilde{P})}{2(\card{\mc{G}_{L(\tilde{P})+1}}+1)} - \Theta(1),
\end{equation}
where we have used that $\card{\mc{G}_{L+1}}=(L+1)^{K^2}$.

From \eqref{eq:achievable} and \eqref{eq:Pbound}
\begin{align}
    \label{eq:clower}
    C(\bm{H},\tilde{P}) 
    & \geq C\bigl(\bm{H},P(L(\tilde{P}),p(\tilde{P}))\bigr) \nonumber\\
    & = K\card{\mc{G}_{L}}\bigl(1-\mc{H}_p(2\varepsilon(p))\bigr)\log(p),
\end{align}
where we have suppressed dependence of $p$ and $L$ on $\tilde{P}$.
By the definition of the $p$-ary entropy function $\mc{H}_p(x)$ in
\eqref{eq:entropy}, we have for any $x\in(0,1)$
\begin{equation*}
    \mc{H}_p(x) \leq x+\mc{H}_2(x)/\log(p).
\end{equation*}
Therefore, \eqref{eq:clower} can be further lower bounded as
\begin{equation*}
    C(\bm{H},\tilde{P}) 
    \geq K\card{\mc{G}_{L}}\bigl((1-2\varepsilon(p))\log(p)
    - \mc{H}_2(2\varepsilon(p))\bigr).
\end{equation*}
Substituting \eqref{eq:pbound},
\begin{equation}
    \label{eq:clower2}
    C(\bm{H},\tilde{P}) 
    \geq \frac{K}{2}(1-2\varepsilon(p))
    \frac{\card{\mc{G}_{L}}}{\card{\mc{G}_{L+1}}+1}\log(\tilde{P}) 
    - \card{\mc{G}_L}\Theta(1).
\end{equation}
From Lemma~\ref{thm:modulation},
\begin{equation*}
    1-2\varepsilon(p(\tilde{P}))
    \geq 1-O\Bigl(\log^{-\tfrac{1}{1+K^2}}(\tilde{P})\Bigr).
\end{equation*}
Moreover,
\begin{align*}
    \frac{\card{\mc{G}_{L(\tilde{P})}}}{\card{\mc{G}_{L(\tilde{P})+1}}+1}
    & = \biggl(\biggl(1+\frac{1}{L(\tilde{P})}\biggr)^{K^2}
    +\frac{1}{(L(\tilde{P}))^{K^2}}\biggr)^{-1} \\
    & \geq 1-O\Bigl(\log^{-\tfrac{1}{1+K^2}}(\tilde{P})\Bigr),
\end{align*}
and
\begin{equation*}
    \card{\mc{G}_{L(\tilde{P})}} 
    = \log^{\tfrac{K^2}{1+K^2}}(\tilde{P}).
\end{equation*}
Combining this with \eqref{eq:clower2} yields
\begin{align*}
    C(\bm{H},\tilde{P})
    & \geq \frac{K}{2}
    \Bigl(1-O\Bigl(\log^{-\tfrac{1}{1+K^2}}(\tilde{P})\Bigr)\Bigr)
    \log(\tilde{P})
    - O\Bigl(\log^{\tfrac{K^2}{1+K^2}}(\tilde{P})\Bigr) \\
    & \geq \frac{K}{2}\log(\tilde{P})
    - O\Bigl(\log^{\tfrac{K^2}{1+K^2}}(\tilde{P})\Bigr).
\end{align*}
Thus, for every $K\geq 2$ and almost every $\bm{H}\in\R^{K\times K}$
there exists a positive constant $c_2=c_2(K,\bm{H})$ such that for all
$\tilde{P}\geq 2$ 
\begin{equation*}
    C(\bm{H},\tilde{P}) 
    \geq \frac{K}{2}\log(\tilde{P})
    - c_2\log^{\tfrac{K^2}{1+K^2}}(\tilde{P}),
\end{equation*}
completing the proof of the theorem. \hfill\IEEEQED

\subsection{Proof of Lemma~\ref{thm:modulation}}
\label{sec:proofslattice3_modulation}

Since modulation involves only a mapping of one symbol at a time, we can
drop the time indices in the following discussion (e.g., we write $x_k$
for $x_k[t]$).

We start by analyzing the power of the transmitted signal $x_k$. Each
$\bar{w}_{k,g}$ takes value in $\{0,\ldots, p-1\}$ and hence has power
$\bar{w}_{k,g}^2 \leq p^2$. Moreover, $\card{\mc{G}_L} = L^{K^2}$,
and each $g\in\mc{G}_L$ satisfies
\begin{equation*}
    \abs{g}
    \leq \bigl(
    \max\bigl\{1,{\textstyle\max_{\tilde{m},\tilde{k}}}\abs{h_{\tilde{m},\tilde{k}}}\bigr\}
    \bigr)^{LK^2}.
\end{equation*}
Thus, the channel input $x_k$ as defined in \eqref{eq:xdef} has power at
most
\begin{align*}
    x_k^2
    & = \Bigl( B {\textstyle\sum_{g\in\mc{G}_L}} \bar{w}_{k,g} g \Bigr)^2 \\
    & \leq B^2 L^{2K^2}p^2 
    \bigl(\max\bigl\{ 1,{\textstyle\max_{\tilde{m},\tilde{k}}}
    \abs{h_{\tilde{m},\tilde{k}}} \bigr\} \bigr)^{2L K^2}.
\end{align*}
Defining the positive constant
\begin{equation*}
    c_4 
    \defeq c_4(K,\bm{H})
    \defeq \bigl(\max\bigl\{1,
    {\textstyle\max_{\tilde{m},\tilde{k}}}\abs{h_{\tilde{m},\tilde{k}}}
    \bigr\}\bigr)^{2K^2},
\end{equation*}
and using the definition of $B$ in \eqref{eq:bdef}, we obtain
\begin{equation*}
    x_k^2 
    \leq c_4^L (Kp)^{2\card{\mc{G}_{L+1}}} L^{2K^2}p^2
\end{equation*}
as required.

We continue by analyzing the probability of demodulation error. Recall
that the received signal
\begin{align*}
    y_m 
    & = \sum_{k=1}^K h_{m,k}x_k  +z_m \\
    & = B\sum_{k=1}^K  \sum_{g\in\mc{G}_L} \bar{w}_{k,g} h_{m,k}g +z_m
\end{align*}
can be rewritten as
\begin{equation}
    \label{eq:yrewrite}
    y_m = B \sum_{g\in\mc{G}_{L+1}}\bar{u}_{m,g}g +z_m,
\end{equation}
with $\bar{u}_{m,g}$ as defined in \eqref{eq:ubardef}.
Receiver $m$ aims to demodulate the functions
$(\bar{u}_{m,g})_{g\in\mc{G}_{L+1}}$ from $y_m$.
We now argue that this is possible with small probability of error.

Consider a different set of linear combinations
$(\bar{u}_{m,g}')_{g\in\mc{G}_{L+1}}\neq(\bar{u}_{m,g})_{g\in\mc{G}_{L+1}}$, 
and compute the difference
\begin{equation}
    \label{eq:udiff}
    \big\lvert{\textstyle\sum_{g\in\mc{G}_{L+1}}} (\bar{u}_{m,g}-\bar{u}'_{m,g})g \big\rvert
    \defeq \big\lvert{\textstyle\sum_{g\in\mc{G}_{L+1}}} q_g g \big\rvert.
\end{equation}
Note that $\bar{u}_{m,g}\in \{0, \ldots, K(p-1)\}$ and hence
$q_g\in\{-K(p-1), \ldots, K(p-1)\}$. Moreover, observe that
$\bar{u}_{m,g}=\bar{u}_{m,g}'=0$ for $g=1$ in any valid set of
equations and almost every $\bm{H}$, so that $q_1 = 0$ and can be ignored. By
Lemma~\ref{thm:groshev2}, for almost every $\bm{H}$ there is a finite
constant $c = c(K, \bm{H})$ such that
\begin{equation*}
    \big\lvert{\textstyle\sum_{g\in\mc{G}_{L+1}, g\neq 1}} q_g g \big\rvert
    \geq c(Kp)^{-\card{\mc{G}_{L+1}}+1/2}
\end{equation*} 
for all $(q_g)_{g\neq 1}\in\{-K(p-1),\ldots, K(p-1)\}^{\card{\mc{G}_{L+1}}-1}$,
$(q_g)_{g\neq 1}\neq \bm{0}$. 

Combined with \eqref{eq:yrewrite} and \eqref{eq:udiff}, this shows that
the minimum distance between any two signal points at receiver $m$ is at
least
\begin{equation*}
    cB(Kp)^{-\card{\mc{G}_{L+1}}+1/2}.
\end{equation*} 
Using the definition of $B$ in \eqref{eq:bdef}, we see that the minimum
distance between the desired set of equations $(\bar{u}_{m,g})_g$ and
any other set of equations $(\bar{u}_{m,g}')_g$ is at least
$c(Kp)^{1/2}$. Therefore, the probability of demodulation error is
upper bounded by
\begin{equation*}
    \Pp\bigl(\cup_g \{\hat{\bar{u}}_{m,g}\neq \bar{u}_{m,g}\} \bigr) \\
    \leq \Pp\bigl(\abs{z_m} > c_5p^{1/2}\bigr)
\end{equation*}
with
\begin{equation*}
    c_5 
    = c_5(K,\bm{H})
    \defeq \frac{cK^{1/2}}{2}.
\end{equation*}
This, in turn, can be upper bounded using the Chernoff bound as
\begin{align*}
    \Pp\bigl(\abs{z_1} > c_5p^{1/2}\bigr)
    \leq \exp\bigl(- \tfrac{1}{2}c_5^2p\bigr),
\end{align*}
concluding the proof of the lemma.  \hfill\IEEEQED

\subsection{Proof of Lemma~\ref{thm:coding}}
\label{sec:proofslattice3_coding}

We derive a lower bound on the rate achievable with linear codes over
the modulated $\F_p$ channel. Recall that each channel encoder uses the
same linear map $\bm{S}$. By \eqref{eq:linear}, the output of the
sub-demodulator $g$ at receiver $m$ reduced modulo $p$ is
\begin{equation*}
    \hat{\bar{\bm{u}}}_{m,g}
    = \bm{S}\bm{u}_{m,g}+\bar{\bm{z}}_{m,g} \pmod{p},
\end{equation*}
with (non-additive) noise $\bar{\bm{z}}_{m,g}$ satisfying 
\begin{equation}
    \label{eq:ind}
    \ind(\bar{z}_{m,g}[t] \neq 0) \leq \ind(\abs{z_m[t]} > c_5p^{1/2}) 
\end{equation}
for all $g\in\mc{G}_{L+1}$, and
\begin{equation}
    \label{eq:eps2}
    \Pp(\abs{z_m[t]} > c_5p^{1/2}) \leq \varepsilon(p) = \varepsilon
\end{equation}
by Lemma~\ref{thm:modulation}. Thus, we only need to analyze the
performance of linear codes over a point-to-point channel with input and
output alphabets $\F_p$ and (non-additive) noise $\bar{z}_{m,g}[t]$. 

By the Gilbert-Varshamov bound (see, e.g., \cite[Theorem 12.3.2, Theorem
12.3.4]{blahut03}), for every $T$, prime number $p$, and $2 \leq d\leq T/2$,
there exists a linear block code of length $T$ over $\F_p$ with rate at
least
\begin{equation*}
    \bigl(1-\mc{H}_p((d-1)/T)\bigr)\log(p),
\end{equation*}
and minimum distance at least $d$. Recall that $0 \leq \varepsilon <
1/4$ by assumption, and fix $\eta\in(0, 1/2-2\varepsilon)$. Choose 
\begin{equation*}
    d = \floor{(2\varepsilon+\eta)T}
    \geq (2\varepsilon+\eta)T-1,
\end{equation*}
so that the linear code can correct up to
\begin{equation*}
    \floor{(d-1)/2} 
    \geq \floor{(\varepsilon+\eta/2)T-1} 
    \geq (\varepsilon+\eta/2)T-2
\end{equation*}
errors. 

Since we use minimum-distance decoding at the receivers, this implies
that we make an error only if the noise has Hamming weight larger than
$(\varepsilon+\eta/2)T-2$, i.e., if
\begin{equation*}
    \max_{g}\sum_{t=1}^T \ind(\bar{z}_{m,g}[t] \neq 0)
    > (\varepsilon+\eta/2)T-2.
\end{equation*}
Using \eqref{eq:ind}, we have
\begin{equation*}
    \Pp\biggl(\max_{g}\sum_{t=1}^T \ind(\bar{z}_{m,g}[t] \neq 0)
    > (\varepsilon+\eta/2)T-2\biggr)
    \leq
    \Pp\biggl(\sum_{t=1}^T \ind(\abs{z_m[t]} \geq c_5p^{1/2})
    > (\varepsilon+\eta/2)T-2\biggr).
\end{equation*}
Since $(z_m[t])_t$ is i.i.d., the weak law of large numbers shows together
with \eqref{eq:eps2} that\footnote{We point out that the law of large
numbers does \emph{not} apply to $\bar{z}_{m,g}[t]$, since this sequence
is dependent on the channel input and therefore, without further
assumptions on those inputs, not i.i.d.}
\begin{equation*}
    \lim_{T\to\infty}
    \Pp\biggl(\max_{g}\sum_{t=1}^T \ind(\bar{z}_{m,g}[t] \neq 0)
    > (\varepsilon+\eta/2)T-2\biggr) = 0.
\end{equation*}
In particular, for $T$ large enough this probability is less than
$\eta/K$, so that with probability at least $1-\eta$ all decoders are able
to decode correctly. The rate of this code is at least
\begin{equation*}
    \bigl(1-\mc{H}_p((d-1)/T)\bigr)\log(p) 
    \geq \bigl(1-\mc{H}_p(2\varepsilon+\eta)\bigr)\log(p),
\end{equation*}
where we have used that $\mc{H}_p(x)$ is increasing in $x$ for $x\leq
1/2$. This proves the lemma.
\hfill\IEEEQED

\subsection{Proof of Lemma~\ref{thm:invertible}}
\label{sec:proofslattice3_invertible}

We need to show that the map from the input to the encoder of the outer code 
\begin{equation*}
    \bigl(\bm{w}_{k,g}: k\in\{1,\ldots, K\}, g\in\mc{G}_L\bigr)
\end{equation*}
to the correct output of the decoder of the outer code
\begin{equation*}
    \bigl(\bm{u}_{m,\tilde{g}}: m\in\{1,\ldots, K\}, \tilde{g}\in\mc{G}_{L+1}\bigr)
\end{equation*}
is injective. The mapping from $\bm{w}_{k,g}$ to $\bar{\bm{w}}_{k,g}$
defined in \eqref{eq:fkgdef} and the mapping from $\bm{u}_{m,\tilde{g}}$
to $\bar{\bm{u}}_{m,\tilde{g}}$ given by \eqref{eq:barbmu2} are both
invertible over their range. Hence, it suffices to prove injectivity of
the map from the input to the modulators
\begin{equation*}
    \bigl(\bar{\bm{w}}_{k,g}: k\in\{1,\ldots, K\}, g\in\mc{G}_L\bigr)
\end{equation*}
to the correct output of the demodulators
\begin{equation*}
    \bigl(\bar{\bm{u}}_{m,\tilde{g}}: m\in\{1,\ldots, K\}, \tilde{g}\in\mc{G}_{L+1}\bigr)
\end{equation*}
over $\F_p$. Finally, since this map is defined at the symbol level, it
suffices to prove injectivity for a single time slot $t$. To simplify
notation, we drop the index $t$ in the following, i.e., we write
$\bar{w}_{k,g}$ and $\bar{u}_{m,\tilde{g}}$ for $\bar{w}_{k,g}[t]$ and
$\bar{u}_{m,\tilde{g}}[t]$. Thus, we need to show that the map from
\begin{equation*}
    \bigl(\bar{w}_{k,g}: k\in\{1,\ldots, K\}, g\in\mc{G}_L\bigr)
\end{equation*}
to
\begin{equation*}
    \bigl(\bar{u}_{m,\tilde{g}}: m\in\{1,\ldots, K\}, \tilde{g}\in\mc{G}_{L+1}\bigr)
\end{equation*}
is invertible over its range. This map is defined in \eqref{eq:ubardef}
as
\begin{equation*}
    \bar{u}_{m,\tilde{g}}[t] 
    = \sum_{k=1}^K \bar{w}_{k,(\tilde{g}/h_{m,k})}[t] \pmod{p},
\end{equation*}
with the convention that $\bar{w}_{k,g}=0$ whenever
$\bar{w}_{k,g}\notin\mc{G}_L$, and taking into account that the
modulator output is reduced modulo $p$.

In the remainder of the proof, we make repeated use of the fact that,
for almost every channel realization $\bm{H}$, every monomial (evaluated
at $\bm{H}$) in $\mc{G}_{L+1}$ can be uniquely factorized into powers
of $(h_{m,k})$, see Remark~\ref{rem:unique} in
Section~\ref{sec:proofspreliminaries}.  We refer to this as the
\emph{unique factorization} property in what follows. For
$g\in\mc{G}_{L+1}$, we use the notation $h_{m,k}^s \mid g$ to denote
that $h_{m,k}^s$ is a factor of $g$ in this unique factorization.

We begin with a small example with $L=K=2$. Recall that the channel gain
between transmitter $k$ and receiver $m$ is denoted by $h_{m,k}$. By
definition, 
\begin{equation*}
    \mc{G}_2 = \bigl\{ 
    h_{1,1}^{s_{1,1}}h_{1,2}^{s_{1,2}}h_{2,1}^{s_{2,1}}h_{2,2}^{s_{2,2}}
    ~: s_{1,1},s_{1,2},s_{2,1},s_{2,2} \in \{0,1\} \bigr\} 
\end{equation*}
and $\vert \mc{G}_2 \vert = 16$ for almost every $\bm{H}$ by unique
factorization.

Consider the received monomial $\tilde{g}\in\mc{G}_3$. If
\begin{equation*}
    h^2_{1,1} \mid \tilde{g}
\end{equation*}
at receiver 1, then $\tilde{g}$ can have originated only from
transmitter 1 by unique factorization. In other words,
$\bar{u}_{1,\tilde{g}} = \bar{w}_{1,(\tilde{g}/h_{1,1})}$ and hence
$\bar{w}_{1,(\tilde{g}/h_{1,1})}$ can be recovered from the received signal. If
\begin{equation*}
    h^2_{2,1} \mid \tilde{g}
\end{equation*}
at receiver 2 than again $\tilde{g}$ can have originated only from
transmitter 1 by unique factorization. Hence $\bar{u}_{2,\tilde{g}} =
\bar{w}_{1,(\tilde{g}/h_{2,1})}$ and 
$\bar{w}_{1,(\tilde{g}/h_{2,1})}$ can be recovered. We can then remove
these messages from all other equations at the receivers.

A similar conclusion can be drawn if $h_{1,2}^2$ is a factor at receiver
1 or respectively $h_{2,2}^2$ a factor at receiver 2. In either case the
monomial originated at transmitter 2, and the corresponding message
$\bar{w}_{2,g}$ can be recovered again by unique factorization and the
terms again be removed from all received signals.

Other parts of the signals may also be identified as only originating
from transmitter 1. For example $h_{1,1} h_{2,1}$ at receiver 1 cannot
be seen as a message from transmitter 2 because $h_{1,2}$ is not a
factor. All such messages can be decoded and removed as was done with
the previous messages, again by unique factorization.

This leaves only the message $\bar{w}_{1,g}$ with $g=h_{1,2} h_{2,2}$
from transmitter 1 and the message $\bar{w}_{2,g}$ with $g=h_{1,1}
h_{2,1}$ from transmitter 2 to be determined. But $h_{1,2} h_{2,2}$
from transmitter 1 is observed at receiver 1 as
\begin{equation*}
    \tilde{g} = g h_{1,1} = h_{1,1} h_{1,2} h_{2,2}
\end{equation*}
which, to have originated from transmitter 2, would be transmitted as 
\begin{equation*}
    \tilde{g}/h_{1,2} = h_{1,1} h_{2,2}.
\end{equation*}
However, this message was already removed in the first round (since it
is observed at receiver 2 as $h_{1,1} h_{2,2}^2$) and so the remaining
signal at receiver 1 must have originated from transmitter 1.  Thus
$\bar{w}_{1,g}$ can be determined. The same can be done for $g = h_{1,1}
h_{2,1}$ from transmitter 2, and the message $\bar{w}_{2,g}$ can be
obtained. This completes the example as all messages $\bar{w}_{k,g}$
for $g \in \mc{G}_2$, $k\in\{1,2\}$ have been determined.

We now extend the above argument to $K=2$ but $L$ arbitrary, proceeding
by induction. We will argue that the messages $\bar{w}_{k,g}, k=\{1,2\},
g \in \mc{G}_L$ are completely determined by $\bar{u}_{m,\tilde{g}}$
where $m\in\{1,2\}$ and $\tilde{g}$ ranges over the possible received
monomials in $\mc{G}_{L+1}$.  The proof is by induction on $L$, and our
earlier argument for $L=2$ anchors the induction. 

Suppose the induction hypothesis holds  for $L -1 \geq 2$. We now show
it holds for $L$ as well. As before, determine and remove from the
received signals all messages $\bar{w}_{1,g}$, such that  $h_{1,1}^{L-1}
\mid g$ or $h_{2,1}^{L-1} \mid g$ at transmitter 1 (using unique
factorization). Do the same, but for messages $\bar{w}_{2,g}$ such that
$h_{1,2}^{L-1} \mid g$ or $h_{2,2}^{L-1} \mid g$ at transmitter 2.

Now, consider $g$ at transmitter 1 such that $h_{1,2}^{L-1} \mid g$. This will
be received as 
\begin{equation*}
    \tilde{g} = g h_{2,1} 
\end{equation*}
at receiver 2, and either $h_{2,2} \mid g$ or it is seen as a monomial component
originating from transmitter 1 immediately by unique factorization. To be from
transmitter 2, the transmit monomial would be
\begin{equation*}
    \tilde{g}/h_{2,2}.
\end{equation*}
But then $h_{1,2}^{L-1} \mid (\tilde{g}/h_{2,2})$ and
therefore the message corresponding to this signal has already been
removed from both receivers. 

The same is true for all messages $\bar{w}_{1,g}$ with $g$ such that
$h_{2,2}^{L-1} \mid g$ at transmitter 1. Moreover the same arguments
apply to transmitter 2 but with $h_{1,1}^{L-1}$ and $h_{2,1}^{L-1}$. It
follows that all factors $\bar{w}_{k,g}$ with the highest exponent in
$g$ being $L-1$ have been determined for both transmitters. The
remaining monomials make up $\mc{G}_{L-1}$ at both transmitters.  Since
the factors involving monomials with highest exponent $L-1$ have been
removed, we may apply the induction hypothesis to complete the
inversion. Thus, for $K=2$ and arbitrary $L\geq 2$, the mapping can be
inverted.

It remains to consider $K \geq 2$ and $L\geq 2$. We will argue that the
factors $\bar{w}_{k,g}$, $k\in\{1,2,\cdots,K\},  g \in \mc{G}_L$ are
completely determined by $\bar{u}_{m,\tilde{g}}$ where
$m=\{1,2,\cdots,K\}$ and $\tilde{g}$ ranges over the possible received
monomials in $\mc{G}_{L+1}$. As earlier, we proceed by induction, but this
time on $K$. The result holds for $K=2$ as we have already demonstrated. 

Suppose then the result holds for $K-1 \geq 2$, and consider the case
with $K$ transmitters and receivers. Fix $L \geq 2$
arbitrarily. For each transmitter $k$, we can once again remove all the
factors $\bar{w}_{k,g}$ whenever
\begin{equation*}
    h_{m,k}^{L-1} \mid g
\end{equation*}
for some $m$.

Now let us fix a receiver, say $\tilde{m}=1$, and a transmitter, say
$\tilde{k} = 2$.  Note this choice is entirely arbitrary as transmitters
are in no sense tied to receivers so we may re-index them to obtain this
case.  Consider a monomial $g$ at transmitter $k$ such that 
\begin{equation*}
    h_{1,2}^{L-1} \mid g.
\end{equation*}
At receiver 2, this is observed as
\begin{equation*}
    \tilde{g} = g h_{2,k}.
\end{equation*}
For such a monomial to be seen at receiver 2 as originating from
transmitter 2, we must have that
\begin{equation*}
    h_{2,2} \mid \tilde{g},
\end{equation*}
since otherwise we can rule out transmitter 2 at receiver 2 by unique
factorization. However, if this is the case, we see that the
corresponding message has already been removed for transmitter 2 as 
\begin{equation*}
    h_{1,2}^{L-1} \mid (\tilde{g}/h_{2,2}).
\end{equation*}
Thus under either outcome this monomial component cannot be seen as
originating from transmitter 2.

Proceed as follows. Consider receivers $m\in\{2,3,\cdots,K\}$ (leaving
out receiver $\tilde{m}=1$) and collect all equations
$\bar{u}_{m,\tilde{g}}$ such that $m \neq 1$ and
\begin{equation*}
    h_{1,2}^{L-1} \mid \tilde{g}
\end{equation*}
using unique factorization. We may consider these as originating only
from transmitters $k\in\{1,3,\cdots,K\}$ (leaving out transmitter
$\tilde{k}=2$) by the argument in the preceding paragraph.  We thus now
have the problem of identifying $\bar{w}_{k,g}$ such that $h_{1,2}^{L-1}
\mid g, k\neq 1$, using the received signals at $m\neq 1$. Denote the
corresponding set of transmit monomials by
\begin{equation*}
    \mc{G}_L^{1,2} 
    \defeq \bigl\{ g\in\mc{G}_L:  h_{1,2}^{L-1} \bigm\vert g \bigr\}.
\end{equation*}
Observe that any power of the channel gains $h_{m,2}, m \neq 1$ and
$h_{1,k}$, $k\neq 2$ may be a factor of the monomials in $\mc{G}^{1,2}_L$. 

To proceed further, note that the monomials in $\mc{G}^{1,2}_L$ can be
partitioned into equivalence classes such that each $g$ in the same
class has the same factors
\begin{equation}
    \label{eq:twofactor}
    h_{1,2}^{L-1}
    \prod_{k\neq 2} h_{1,k}^{s_{1,k}}
    \prod_{m\neq 1} h_{m,2}^{s_{m,2}}
\end{equation}
in their unique factorization for some fixed $0 \leq s_{1,k} \leq L-1, 0
\leq s_{m,2} \leq L-1$. We call \eqref{eq:twofactor} the
\emph{(1,2)-factor} of $g$. We may also partition the receive monomials
according to their (1,2)-factor. Denote by
\begin{equation*}
    \mc{G}_L^{1,2}\bigl((s_{1,k})_{k\neq 2},(s_{m,2})_{m\neq 1}\bigr)\subset \mc{G}_L^{1,2}
\end{equation*}
the equivalence class with (1,2)-factor 
\begin{equation*}
    h_{1,2}^{L-1}
    \prod_{k\neq 2} h_{1,k}^{s_{1,k}} \prod_{m \neq 1} h_{m,2}^{s_{m,2}}.
\end{equation*}

Fix a class $(s_{1,k})_{k\neq 2}, (s_{m,2})_{m\neq 1}$, 
and consider the messages
\begin{equation}
    \label{eq:messages}
    \Bigl(\bar{w}_{k,g}: k\neq 2, 
    g\in\mc{G}_L^{1,2}\bigl((s_{1,k})_{k\neq 2}, (s_{m,2})_{m\neq 1}\bigr)\Bigr)
\end{equation}
and the equations
\begin{equation}
    \label{eq:equations}
    \Bigl(\bar{u}_{m,\tilde{g}}: 
    m\neq 1, \tilde{g}\in\mc{G}_{L+1}^{1,2}\bigl((s_{1,k})_{k\neq 2}, 
    (s_{m,2})_{m\neq 1}\bigr)\Bigr).
\end{equation}
Recall that we have removed all messages $\bar{w}_{2,g}$ from the
equations \eqref{eq:equations}. Observe that any equation
$\bar{u}_{m,\tilde{g}}$ in \eqref{eq:equations} is then solely a function of the
messages in \eqref{eq:messages}.

Divide out the common (1,2)-factor from the transmit and receive monomials
in \eqref{eq:messages} and \eqref{eq:equations}. This results in a set
of messages and equations with monomials in the channel coefficients
$h_{m,k}, k \neq 2, m \neq 1$ with $K-1$ transmitters and $K-1$
receivers. By our induction hypothesis, we may invert to obtain all
$\bar{w}_{k,g}, g \in \mc{G}^{1,2}_L$, working with each (1,2)-factor
class in turn.

However, the choice of $\tilde{k}=1, \tilde{m}=2$ plays no special role in the above
argument as we have already explained, so that we may recover
$\bar{w}_{k,g}$ for all monomials 
\begin{equation*}
    \bigl\{ g\in\mc{G}_L : h_{\tilde{m},\tilde{k}}^{L-1} \bigm\vert g \bigr\}
\end{equation*}
and any choice $\tilde{k},\tilde{m}\in\{1,2,\cdots,K\}$. Removing these
decoded messages from the received equations, we have reduced the
monomials to have exponent no higher than $L-2$. Hence, we may proceed
iteratively, reducing the order of $L$ by one in each iteration. Thus,
$\bar{w}_{k,g}$ can be recovered for all $k\in\{1,2,\cdots,K\}, g \in
\mc{G}_L$, that is, the mapping is invertible over its range.
\hfill\IEEEQED

\section{Conclusion}
\label{sec:conclusion}

We considered the asymptotic behavior of compute-and-forward over a
section of a relay network with $K$ transmitters and $K$ relays. We
showed that the lattice implementation of compute-and-forward proposed
by Nazer and Gastpar in \cite{nazer11} achieves at most $2/(1+1/K)\leq
2$ degrees of freedom.  Thus, the asymptotic behavior of the lattice
scheme is very different from the MIMO upper bound resulting from
allowing full cooperation among transmitters and among relays and
achieving $K$ degrees of freedom.  We then argued that this gap is not
fundamental to the compute-and-forward approach in general, but rather
due to the lattice implementation in \cite{nazer11}. To this end, we
proposed and analyzed a different implementation of compute-and-forward
and showed that it achieves $K$ degrees of freedom. Thus, at least in
terms of degrees of freedom, compute-and-forward can achieve the same
asymptotic rates as if full cooperation among transmitters and among
relays were permitted.

\section{Acknowledgments}

The authors would like to thank the anonymous reviewers for their
careful reading of the manuscript and their thoughtful comments.

\appendices

\section{Change of Measure in the Proof of Theorem~\ref{thm:lattice1}}
\label{sec:measure}

Here we show that if \eqref{eq:lattice1_3b} in
Section~\ref{sec:proofslattice1} holds for almost all
$\tilde{\bm{h}}\in\R^{K-1}$, then it also holds for almost all
$\bm{h}\in\R^K$. In the following discussion, we use the notation
$\mu_K$ to denote Lebesgue measure over $\R^K$.  Let $B\subset\R^{K-1}$
be a set of vectors $\tilde{\bm{h}}\in\R^{K-1}$ of measure zero, i.e.,
\begin{equation*}
    \mu_{K-1}(B)=0.
\end{equation*}
Let $D\subset\R^K$ be the set of vectors
$\bm{h}\in\R^K$ such that
\begin{equation*}
    \frac{1}{\norm{\bm{h}}}
    \begin{pmatrix}
        h_1 & \cdots & h_{K-1}
    \end{pmatrix}
    \in B.
\end{equation*}

We want to show that $D$ has also measure zero, i.e., $\mu_{K}(D)=0$. We
have
\begin{align*}
    \mu_K(D) 
    & = \int_{\bm{h}\in\R^K} \ind_D(\bm{h})d\bm{h} \\
    & = \int_{\bm{h}\in\R^K} 
    \ind_B\Bigl(\tfrac{1}{\norm{\bm{h}}}
    \begin{pmatrix}
        h_1 & \cdots & h_{K-1} 
    \end{pmatrix}
    \Bigr)d\bm{h}.
\end{align*}
Making the change of variables
\begin{align*}
    \tilde{h}_k & \defeq \frac{h_k}{\norm{\bm{h}}}, \quad \text{for $k\in\{1,\ldots, K-1\}$}, \\
    s & \defeq \norm{\bm{h}},
\end{align*}
and using the nonnegativity of $\ind_B$ together with Fubini's theorem,
we can rewrite this as
\begin{align*}
    \mu_K(D)
    & = \int_{s=0}^\infty s^{K-1} \int_{\tilde{\bm{h}}\in\R^{K-1}:\norm{\tilde{\bm{h}}}\leq 1}
    \ind_B(\tilde{\bm{h}})(1-\norm{\tilde{\bm{h}}}^2)^{-1/2}d\tilde{\bm{h}}ds.
\end{align*}

Now, 
\begin{align*}
    \int_{\tilde{\bm{h}}\in\R^{K-1}:\norm{\tilde{\bm{h}}}\leq 1} & 
    \ind_B(\tilde{\bm{h}})(1-\norm{\tilde{\bm{h}}}^2)^{-1/2}d\tilde{\bm{h}} \\
    & \leq
    \int_{\tilde{\bm{h}}\in\R^{K-1}:\norm{\tilde{\bm{h}}}\leq \sqrt{1-\varepsilon^2}}
    \ind_B(\tilde{\bm{h}})\varepsilon^{-1}d\tilde{\bm{h}} 
    + \int_{\tilde{\bm{h}}\in\R^{K-1}:
    \sqrt{1-\varepsilon^2} < \norm{\tilde{\bm{h}}}\leq 1}
    (1-\norm{\tilde{\bm{h}}}^2)^{-1/2}d\tilde{\bm{h}}  \\
    & \leq \varepsilon^{-1}\mu_{K-1}(B) 
    + 2\frac{\pi^{(K-1)/2}}{\Gamma((K-1)/2)}
    \int_{\sqrt{1-\varepsilon^2} < \tilde{s} \leq 1} 
    (1-\tilde{s}^2)^{-1/2}d\tilde{s} \\
    & = 2\frac{\pi^{(K-1)/2}}{\Gamma((K-1)/2)}
    \Bigl(\pi/2-\arcsin\bigl(\sqrt{1-\varepsilon^2}\bigr)\Bigr)
\end{align*}
for every $\varepsilon>0$, and where $\Gamma(\cdot)$ denotes 
the Gamma function.  Letting $\varepsilon\to 0$, we obtain
\begin{equation*}
    \int_{\tilde{\bm{h}}\in\R^{K-1}:\norm{\tilde{\bm{h}}}\leq 1} 
    \ind_B(\tilde{\bm{h}})(1-\norm{\tilde{\bm{h}}}^2)^{-1/2}d\tilde{\bm{h}}  
    = 0,
\end{equation*}
and hence
\begin{equation*}
    \mu_K(D) = 0. 
\end{equation*}
This shows that \eqref{eq:lattice1_3b} holds also for almost every
$\bm{H}\in\R^{K\times K}$.

\bibliography{journal_abbr,diophantine}

\end{document}